\documentclass[
superscriptaddress,
twocolumn,
nofootinbib,
 amsmath,amssymb,
 aps,
prd,
]{revtex4-2}

\usepackage[dvipsnames]{xcolor}
\usepackage{graphicx}
\usepackage{dcolumn}
\usepackage{bm}
\usepackage{hyperref}
\usepackage[mathlines]{lineno}
\usepackage{mathrsfs}
\usepackage[normalem]{ulem}
\usepackage{blindtext}

\usepackage{orcidlink}

\def\Q{\mathcal{Q}}

\def\k{{\bf k}}
\def\CII{[C${\rm \textsc{ii}}$]}
\def\CI{[C${\rm \textsc{i}}$]}

\def\OIII{[O${\rm \textsc{iii}}$]}
\def\NII{[N${\rm \textsc{ii}}$]}

\def\B19{\texttt{B19}}
\def\HI{H${\rm \textsc{i}}$}
\def\mpci{{\rm Mpc}^{-1}}
\def\z{{\mathcal{Z}}}

\begin{document}

\preprint{APS/123-QED}

\title{Clues from $\mathcal{Q}$ -- 
A null test designed for line intensity mapping cross-correlation studies}

\author{Debanjan Sarkar\,\orcidlink{0000-0001-5763-2541}}
 \email{debanjan.sarkar@mcgill.ca}
\affiliation{Department of Physics and Trottier Space Institute, McGill University, QC H3A 2T8,
Canada}%
 \affiliation{Ciela—Montreal Institute for Astrophysical Data Analysis and Machine Learning, QC H2V
0B3, Canada}

\author{Ella Iles\,\orcidlink{0009-0005-2877-7692}}
 \altaffiliation[]{}
 \email{ella.iles@mail.mcgill.ca}
 \affiliation{Department of Physics and Trottier Space Institute, McGill University, QC H3A 2T8,
Canada}

\author{Adrian Liu\,\orcidlink{0000-0001-6876-0928}}%
 \email{adrian.liu2@mcgill.ca}
\affiliation{Department of Physics and Trottier Space Institute, McGill University, QC H3A 2T8,
Canada}
 \affiliation{Ciela—Montreal Institute for Astrophysical Data Analysis and Machine Learning, QC H2V
0B3, Canada} 


\begin{abstract}

Estimating the auto power spectrum of cosmological tracers from line-intensity mapping (LIM) data is often limited by instrumental noise, residual foregrounds, and systematics. Cross-power spectra between multiple lines offer a robust alternative, mitigating noise bias and systematics. However, inferring the auto spectrum from cross-correlations relies on two key assumptions: that all tracers are linearly biased with respect to the matter density field, and that they are strongly mutually correlated. In this work, we introduce a new diagnostic statistic, \( \mathcal{Q} \), which serves as a data-driven null test of these assumptions. 
Constructed from combinations of cross-spectra between four distinct spectral lines, \( \mathcal{Q} \) identifies regimes where cross-spectrum-based auto-spectrum reconstruction is unbiased. We validate its behavior using both analytic toy models and simulations of LIM observables, including star formation lines ([C$\textsc{ii}$], [N$\textsc{ii}$], [C$\textsc{i}$],[O$\textsc{iii}$]) and the 21-cm signal. We explore a range of redshifts and instrumental configurations, incorporating noise from representative surveys. Our results demonstrate that the criterion \( \mathcal{Q} \approx 1 \) reliably selects the modes
where cross-spectrum estimators are valid, while significant deviations are an indicator that the key assumptions have been violated. The \( \mathcal{Q} \) diagnostic thus provides a simple yet powerful data-driven consistency check for multi-tracer LIM analyses.

\end{abstract}

\maketitle

\section{Introduction}

\begin{figure}[h!]
    \centering
    \includegraphics[width=1\linewidth]{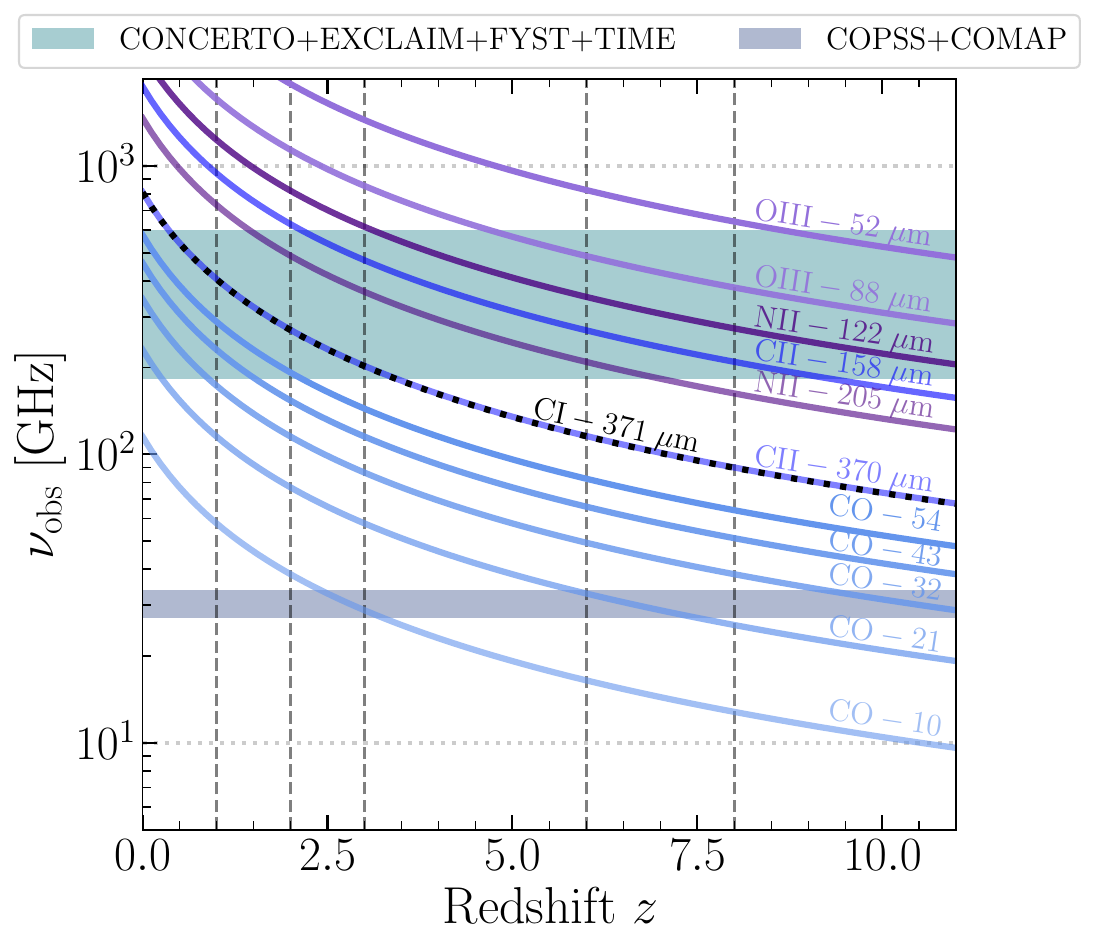}
    \caption{
    The observed frequency $\nu_{\rm obs}$ of selected star formation lines as a function of redshift $z$. The shaded regions denote the frequency coverage of various 
    LIM experiments, as shown in the top legend.  
    The five vertical dashed lines correspond to $z=1,2,3, 6$ and $8$.
    At a given redshift, solid lines passing through a given shaded region will be observed by 
    that corresponding experiment. The horizontal dotted straight lines represent the $10-1000$ GHz frequency coverage of the hypothetical Super-LIM experiment considered in this paper.}
    \label{fig:expts}
\end{figure}

Mapping our Universe by measuring the intensities of characteristic spectral line emissions has emerged as a promising approach for probing the structure of the early universe. This technique, also known as Line Intensity Mapping (LIM), has the potential to make large maps of large-scale structures in a relatively short amount of time compared to traditional optical galaxy surveys
\cite{Kovetz:2017agg,Kovetz:2019uss,Bernal:2019jdo,Bernal:2022jap}. One of the prime examples is LIM using the 21-cm line, which is emitted from spin-flip transitions in neutral hydrogen (\HI) \cite{Bharadwaj:2000av, Furlanetto:2006jb, 2012RPPh...75h6901P}. A number of existing and upcoming surveys of the 21 cm line, such as the Canadian Hydrogen Intensity Mapping Experiment (CHIME) \cite{CHIME:2022kvg,CHIME:2022dwe}, the Canadian Hydrogen Observatory and Radio transient Detector (CHORD)~\cite{2019clrp.2020...28V}, the Hydrogen Epoch of Reionization Array (HERA) ~\cite{HERA:2021bsv,HERA:2021noe}, the Low-Frequency Array (LOFAR)~\cite{van_Haarlem_2013}, the Murchison Widefield Array (MWA)~\cite{Bowman_2013}, and the Square Kilometre Array (SKA)\cite{Koopmans:2015sua} are poised to make detections of spatial 21 cm fluctuations over a large range of redshifts, potentially covering the cosmic dawn to the present epoch \cite{DeBoer_2017,Furlanetto:2006jb, Pritchard:2011xb, Sarkar:2016lvb, Sarkar:2022dvl}. 
While 21 cm intensity mapping has been a focus for some time, there is now growing interest in experiments targeting emission lines for LIM from star-forming regions over a variety of redshifts,~\cite{Gong_2017,Yang_2021,Karkare:2022bai,Libanore:2022ntl}. Several current and upcoming experiments are targeting various star-formation lines
\footnote{\href{https://lambda.gsfc.nasa.gov/product/expt/lim\_experiments.html}{https://lambda.gsfc.nasa.gov/product/expt/lim\_experiments.html}}
such as Tomographic Ionized-Carbon Mapping Experiment (TIME), Fred Young Submillimeter Telescope (FYST), EXperiment for Cryogenic Large-Aperture Intensity Mapping (EXCLAIM), CarbON \CII\ line in post-rEionization and ReionizaTiOn epoch (CONCERTO) for \CII\ ~\cite{Roy:2024dmt,Gong:2011mf, 10.1117/12.2057207,Uzgil:2014pga,Yue:2015sua,refId0,Padmanabhan:2018yul,Chung:2018szp,Pullen:2017ogs,Yang:2019eoj,CONCERTO:2020ahk,Ade:2019ril, 10.1117/12.2561069, 10.1117/12.2630054, Bethermin:2022lmd, CCAT-Prime_Collaboration_2023, VanCuyck:2023uli}, CO Mapping Array Pathfinder (COMAP) and CO Power Spectrum Survey (COPSS) for the CO rotational lines~\cite{Lidz_2011,Carilli_2011,Keating:2016pka,Keating:2015qva,Keating:2020wlx,Karkare:2022bai,Cleary:2021dsp,Stutzer:2024rps,COMAP:2024iqp,COMAP:2021nrp,COMAP:2021sqw,COMAP:2024pkb}, H$\alpha$/H$\beta$~\cite{Gong_2017}, $[O\:\textsc{ii}/\textsc{iii}]$~\cite{Padmanabhan:2021tjr}, and Spectro-Photometer for the History of the Universe, Epoch of Reionization, and Ices Explorer (SPHEREx) for Ly$\alpha$~\cite{Silva_2013,Pullen:2013dir,Croft:2018rwv,Mas-Ribas:2020wkz, Kakuma:2019afo,Renard:2020mfg,Kikuchihara_2022,SPHEREx:2014bgr}. These lines are relatively bright and can be observed over a broad range of redshifts. We illustrate this in Figure~\ref{fig:expts}, where we show the (redshifted) observed frequencies of selected star formation lines along with the frequency coverage of various LIM experiments. Each of these lines will be a powerful probe of astrophysics~\cite{Sun:2020mco,Mirocha_2022,Parsons:2021qyw,Bernal:2019gfq,Horlaville:2023ceh} and cosmology~\cite{Fonseca:2016qqw,Bernal:2019jdo,Schaan:2021hhy,MoradinezhadDizgah:2018zrs,MoradinezhadDizgah:2021upg,MoradinezhadDizgah:2021dei,Karkare:2022bai} on their own. It is clear that in the near future, we will be in a regime where a given redshift can be surveyed by many different lines, opening the potential for multi-tracer techniques.

Having a multi-line view of the same cosmological volume allows for a complementary view of relevant physical processes. For instance, the 21 cm line probes neutral hydrogen regardless of its origin. This means that during (and before) reionization, it mostly originates from neutral gas in the intergalactic medium (IGM) \cite{Madau:1996cs,Wise:2019qtq,Choudhury:2022rlm}, 
whereas after reionization it traces self-shielded pockets of hydrogen found within galaxies
\cite{Bharadwaj:2000av,Villaescusa-Navarro:2018vsg,Kamran:2024xob,Pourtsidou:2022gsb}. The star formation lines typically originate from highly ionized star-forming regions, with the precise details dependent on the local thermodynamic and radiative environment. A truly comprehensive view of galaxy formation and cosmology will likely require multiple complementary surveys with multiple lines~\cite{Pullen:2012su,Sun_2019,Schaan:2021gzb,Anderson:2022svu,Maniyar:2021arp, Sato-Polito:2022wiq, Sun:2022ucx, Mas-Ribas:2022jok,Roy:2023pei,Roy:2023cpx, LujanNiemeyer:2023vfz,Fronenberg:2023juh,Padmanabhan:2022elb}.

One way to statistically exploit the aforementioned complementarity is to perform cross-correlations, such as by forming the cross power spectrum. In addition to being a convenient summary statistic that probes the physics of two different maps, a cross power spectrum has the added benefit of suppressing various contaminations in a signal.
As an example of a contaminant, most intensity mapping surveys need to deal with foregrounds~\cite{Liu_2011}. These might include continuum foregrounds such as synchrotron radiation and dust, as well as interloper foregrounds where an untargeted spectral line is redshifted into the same observing band as a targeted line of interest~\cite{Wang:2005zj,Santos:2004ju, 2008MNRAS.389.1319J}. Additionally, instrumental systematics pose significant challenges, adding further complexity to observations and data analysis~\cite{Lanman_2019}. Cross-correlations provide a potential avenue forward amidst such systematics, since these contaminants will---on average---cross correlate away between two independent surveys with independent systematics. Although the cross-correlations do not get rid of the increased \emph{variance} (and therefore error bars) due to the contaminants \cite{Fronenberg:2024olu}, they provide an attractive way to obtain unbiased measurements.

In fact, cross-correlations can be taken one step further and can be used to extract 
the auto-correlation power spectrum of a spectral line solely from cross-power spectra between three spectral lines. This formalism was first proposed in Ref.~\cite{B19,Beane:2018pmx}, 
and later generalized in Ref.~\cite{LisaPaper}, where
the authors show that the large-scale 21-cm power spectrum from the late cosmic dawn (CD) and reionization epoch (EoR)
can be extracted using only cross-power spectra between the 21 cm fluctuations and 
two separate line-intensity mapping data cubes, like \CII\ and \OIII.
We refer to this as \B19 estimator after Ref.~\cite{B19}. This estimator relies on the following assumptions:
(i) all the lines trace the underlying matter distribution following a linear biasing scheme, and
(ii) the lines are perfectly correlated. These ensure that by taking suitable combinations of 
cross-correlation power spectra of three lines, we can recover the auto power spectrum. 
These assumptions, however, are true only for a limited
range of spatial wavenumbers $k$, beyond which the \B19 estimator is not reliable. 
Estimating the power spectrum 
with this estimator alone does not provide any information on this $k$ range.
One possibility is to compare the estimation from \B19 with the pure auto power spectrum estimator and the $k$ range would simply be the range over which 
these two estimations match. 
However, the premise here is that because of contaminant biases, the auto power spectrum is observationally unavailable in the first place. Of course, one could resort to theory and simulations to test the \B19 estimator's robustness. But while simulations may be necessary before deploying B19, they may not be sufficient given modeling uncertainties such as the complicated radiative transfer effects of some difficult-to-model spectral lines.

In this paper, we introduce a new diagnostic quantity \( \mathcal{Q} \) that enables a \emph{data-driven null test} for evaluating the two fundamental assumptions---linear biasing and strong mutual correlation---underlying the \B19\ formalism. This statistic is constructed using a specific combination of cross-correlations among four distinct spectral lines that can be jointly observed at a given redshift. As illustrated in Figure~\ref{fig:expts}, such line combinations are accessible through a combination of multiple current and upcoming LIM experiments, making \( \mathcal{Q} \) a practically applicable consistency check. 
In this paper, we discuss in detail the formalism of the \(\Q\) statistic and test 
its validity for a number of realistic scenarios, including the effects of instrumental noise. 
We demonstrate how \(\Q\) can be used to evaluate the reliability of the \B19 estimator.

The paper is organized as follows. 
In Section~\ref{sec:formalism}, we review the theoretical foundation of the \B19\ estimator and formally derive the \( \mathcal{Q} \) statistic from combinations of cross-power spectra involving four distinct spectral lines. We then validate the behavior of \( \mathcal{Q} \) using controlled toy models with known correlation and bias structures. In Section~\ref{sec:sf_lines}, we assess the performance of \( \mathcal{Q} \) using astrophysically motivated simulations of star formation lines, exploring its dependence on physical parameters, such as minimal halo mass and instrumental noise. We provide a classification framework for the estimator performance and show scenarios for every outcome. Section~\ref{sec:with_21cm} extends the analysis to configurations involving the 21-cm line after the Epoch of Reionization, evaluating the estimator's sensitivity to changes in tracer properties. Finally, we conclude in Section~\ref{sec:summary} with a summary of our findings and a discussion of future applications of the \( \mathcal{Q} \) diagnostic in multi-tracer LIM analyses.

\section{The $\mathcal{Q}$-estimator}
\label{sec:formalism}

In this section, we define a data-driven method based on an estimator $\Q$ that allows one test the applicability of the \B19
power spectrum estimator. 
The \texttt{B19} estimator \cite{B19,Beane:2018pmx,LisaPaper} proposes recovering the auto spectrum of a spectral line by finding the cross-correlation spectrum between three spectral lines and then performing a division to recover an estimator $\hat{P}_{aa}$ of the auto spectrum given by
\begin{equation}
\label{eq:B19}
     \hat{P}_{aa} = \frac{P_{ab}P_{ac}}{P_{bc}}.
\end{equation}
Here $a$, $b$, and $c$ refer to different spectral lines and $P_{ab}$, $P_{ac}$, \dots
refer to the cross power spectra between the different pairs of lines.
The success of the \texttt{B19} estimator mainly depends on three assumptions,
(i) the systematics between the different spectral lines are uncorrelated,  (ii) the spectral lines trace the underlying matter distribution linearly, 
and (iii) spectral lines are highly correlated with each other. 
If these assumptions are satisfied, the \texttt{B19} estimator is expected to provide unbiased 
estimation of the auto power spectrum $P_{a}$.
The uncorrelated systematics ensures that when we cross-correlate two different lines, 
the systematics get filtered out and we obtain clean cross-correlation spectra. 
The linear biasing model states that a spectral line fluctuation field $\delta_a(\k)$ follows the 
underlying matter overdensity field $\delta_m(\k)$ as 
\begin{equation}
    \delta_a (\k) = \beta_a(k) \delta_m(\k)\,,
\end{equation}
where we have written the fields in Fourier $\mathbf{k}$-space and $\beta_a(k)$ is the linear bias parameter which 
in general can be $k$-dependent. Now if we define the cross power spectrum between two fields 
$a$ and $b$ as 
\begin{equation}
     \langle \delta_a (\k) \delta^{\ast}_b (\k^{\prime}) \rangle = (2\pi)^3  \delta_{\rm D}(\k-\k^{\prime}) P_{ab}(k),
\end{equation}
where $\delta_{\rm D}$ is the Dirac delta function and the angular bracket refers to the ensemble average, then we can express the cross power spectrum 
in principle as
\begin{equation}
    P_{ab}=\beta_a \beta_b P_m\,,
    \label{eq:cross-corr}
\end{equation}
where $P_m$ is the underlying matter power spectrum. In practice it is more accurate to say
\begin{equation}
    P_{ab}=\beta_a \beta_b r_{ab} P_m,
    \label{eq:cross-corr2}
\end{equation}
where $r_{ab}$ is a cross-correlation 
coefficient between the two lines that ranges between $\pm1$. Note that we have dropped the $k$-dependence for brevity (and will continue to do so from here onward).
Using this expression, the \texttt{B19} estimator can be rewritten as 
\begin{equation}
    \hat{P}_{aa} = \left(\frac{r_{ab}r_{ac}}{r_{bc}}\right)
    \frac{(\beta_a \beta_b P_m)(\beta_a \beta_c  P_m)}{(\beta_b \beta_c P_m)}\,.
    \label{eq:b19}
\end{equation}
Under our current assumption of linear biasing, the bias factors for the two lines 
$b$ and $c$ cancel. In addition, one expects that on large scales, the ratio of the cross-correlation coefficients also
becomes close to unity, i.e. $r_{ab}r_{ac}/r_{bc}\rightarrow1$. Thus, 
from \texttt{B19} estimator we recover
\begin{equation}
    \hat{P}_{aa} = \beta^2_a P_m\,,
\end{equation}
which is a (statistically) unbiased estimation of the true auto power spectrum $P_a$ of line $a$. 

Inspired by the cancellations seen above, we consider four different lines and design an estimator in the
following way. We compute cross power spectra of the lines and arrange those in such a way that the matter power spectrum and the bias terms get canceled, i.e.,
\begin{equation}
    \frac{P_{ab}P_{cd}}{P_{ac}P_{bd}} = \frac{r_{ab}r_{cd}}{r_{ac}r_{bd}}\,.
\end{equation}
In general, this is true for 
other permutaions of $a,b,c$ and $d$
as long as the bias terms cancel.
At large scales, $r_{ab}r_{cd}/r_{ac}r_{bd} \rightarrow 1$ and the estimator converges to
$\rightarrow 1$. This, however, may not be true in scenarios where the linear biasing does not hold, or the nature of cross-correlations
are different between the lines. Considering these, we define an estimator,
\begin{equation}
    \mathcal{Q}_{abcd} = \frac{P_{ab}P_{cd}}{P_{ac}P_{bd}}\,,
\end{equation}
(or simply \(\Q\)-estimator)
which can be considered as a null test where deviations from $\mathcal{Q}=1$ imply that the fields being considered do not trace each other well enough. In the rest of the paper, we pass the estimator
through a number of realistic tests to check whether it deviates from unity, and if so, 
what we can learn from it.  In order to test the \(\Q\) estimator,  
we will use realistic simulations of spectral line fields
that will be observed in a number of existing and upcoming experiments. Prior to that, we build intuition by testing the estimator's robustness on simple fields where we can control the correlations between the fields.
For the rest of this section we will discuss two simple scenarios.

\begin{figure}
    \centering
    \includegraphics[width=0.9\linewidth]{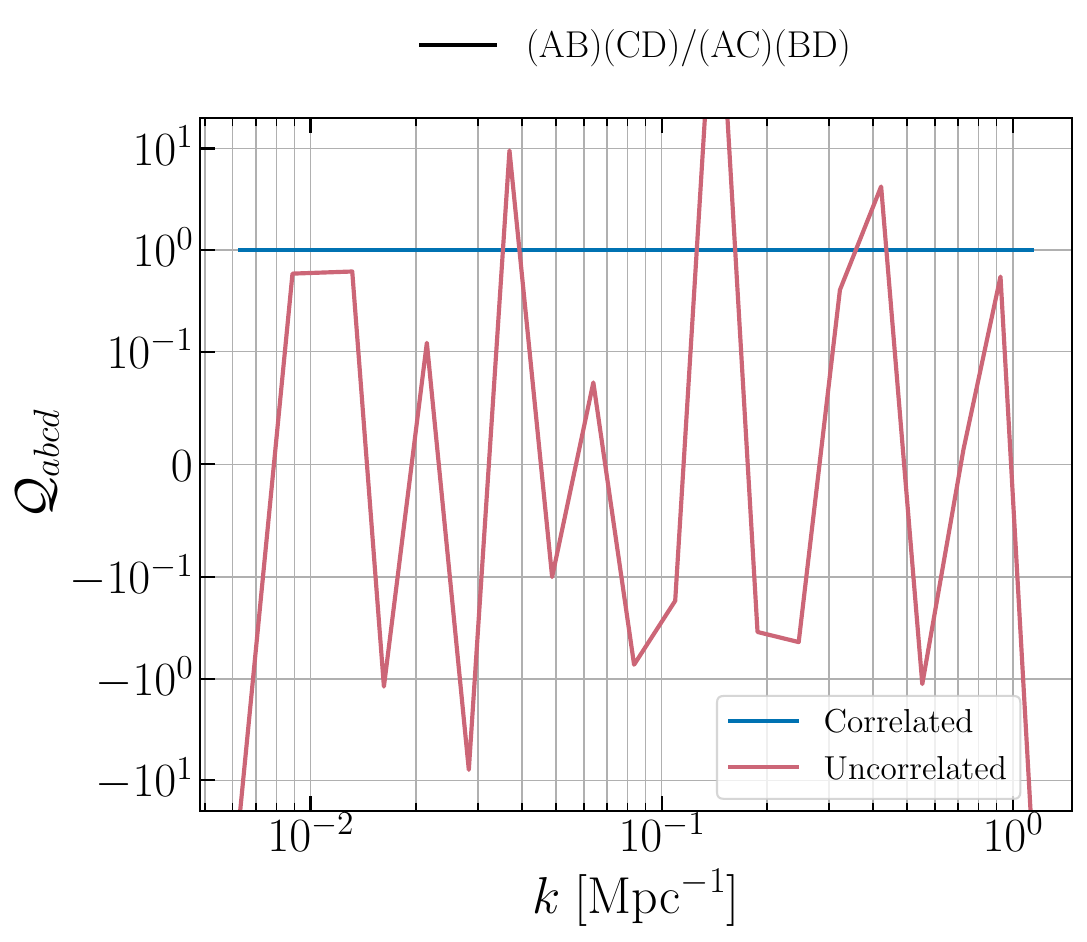}
    \caption{The $\mathcal{Q}$ estimator as a function of spatial wavenumber $k$ for two toy models. The blue horizontal line represents $\mathcal{Q}$ for four perfectly correlated realizations of density 
    fields generated from a realization of a matter density field
    at $z=6$, while the red fluctuating line shows the result for four independent random Gaussian fields that are completely uncorrelated by construction. The \(\Q\) statistic is computed using 
    a single combination of the cross-power spectra $(P_{\rm AB}P_{\rm CD})/(P_{\rm AC}P_{\rm BD})$; 
    it is equal to unity at all $k$ for the perfectly correlated 
    fields, which is expected. However, this is not true for the uncorrelated fields, and 
    we see large fluctuations in \(\Q\) due to the presence of zeroes in the denominator. 
    }
    \label{fig:Q_trivial}
\end{figure}

In the first scenario, we take a realization of a matter density field at $z=6$, 
generated on a $(256)^3$ grid within a volume of $(1000\,{\rm Mpc})^3$. Starting from the linear power spectrum at $z=6$ based on \textit{Planck} 2018 parameters \cite{Planck:2018vyg}, we use the \texttt{nbodykit}\footnote{\href{https://nbodykit.readthedocs.io/en/latest/}{https://nbodykit.readthedocs.io/en/latest/}} package \cite{Hand:2017pqn} to generate a Gaussian random field representing the matter density field.
We denote this field as $A$. 
From $A$, we obtain three linearly biased fields $B$, $C$ and $D$ by multiplying field $A$ with 
constant bias factor $\beta_B=2$, $\beta_C=4$, $\beta_D=6$. Therefore, fields $B$, $C$ and $D$
are perfectly correlated with field $A$, as they are only scaled versions of $A$.  
In Figure~\ref{fig:Q_trivial}, we show the \(\Q\) estimator computed using these four 
correlated fields with one combination
of the cross-correlation spectra. 
We find that $\Q=1$ at all $k$, which is expected for perfectly correlated fields. 
Our choice of matter density field here is completely arbitrary; however, it will  
help us to connect with the results discussed in the later sections. We note that 
$\Q=1$ is true even for different signs of the bias factors. It is also important to notice that in Figure ~\ref{fig:Q_trivial}, the blue line is perfectly flat, with no scatter. This illustrates how $\Q$ is a quantity that overcomes cosmic variance, because all the randomness of the fields cancel out, as is often the case with ratio statistics \cite{2009JCAP...10..007M}.

In the second scenario, we generate independent realizations of Gaussian random fields on 
$(256)^3$ grids spanning over a volume of $(1000\,{\rm Mpc})^3$. The choice of the
volume is to be consistent with the previous scenario. We take four of these independent boxes
and estimate \(\Q\). The boxes are uncorrelated by construction, and they should 
have zero correlation. Therefore, the \(\Q\) estimator can be undefined in this case as it contains ratios
of zeroes. In Figure~\ref{fig:Q_trivial}, this is shown in red, and we see large fluctuations 
around zero due to the presence of zeroes in the denominator. Therefore when fields don't correlate, these large fluctuations can cause non-trivial error statistics \citep{LisaPaper}.

In practice, one expects to see behavior somewhere between the blue and red curves of Figure~\ref{fig:Q_trivial}. Cosmological line intensity fields are in general correlated to each other as 
they trace the same matter density field, suggesting something along the lines of the blue curve. Their degree of correlation, however, may vary, and based on that, we may obtain values of \(\Q\) that are non-unity. Additionally, in real observations there will be some length scales that are completely dominated by noise and the cross-correlations yield zero at those scales. The \(\Q\) estimator would then behave more like the red curve. We shall discuss the exact mixture of these behaviors in the following sections.

The order in which fields appear in the $\Q$ estimator is arbitrary but can influence its behavior. For example, a cross power spectrum close to zero makes the $\Q$ estimator diverge if it is in the denominator, but not if it is in the numerator. It is therefore important to consider all combinations of fields when investigating this estimator. Here we chose three different combinations of the cross power spectrum to compute the $\Q$ estimator. For any four lines a, b, c and d, these are denoted by 
\begin{equation}
    \Q_1 \equiv \frac{P_{\rm ab}P_{\rm cd}}{P_{\rm ac}P_{\rm bd}}; \quad
    \Q_2 \equiv \frac{P_{\rm ad}P_{\rm bd}}{P_{\rm ab}P_{\rm cd}}; \quad
    \Q_3 \equiv \frac{P_{\rm ac}P_{\rm bd}}{P_{\rm ad}P_{\rm bc}}\,.
    \label{eq:Qi_defs}
\end{equation}
Wherever we use these symbols, we stick to this definition. 
For the \B19 estimated power spectrum, we also use three different combinations of the cross power spectra.

\section{Star Formation Lines}
\label{sec:sf_lines}

\begin{table*}[t]
\centering
\begin{tabular}{cccccccc}
\hline\hline
Redshift & $M_0$ (${\rm M_{\odot}}$) & $M_2$ (${\rm M_{\odot}}$) & $M_3$ (${\rm M_{\odot}}$) & $\mathfrak{a}$ & $\mathfrak{b}$ & $\mathfrak{c}$ \\
\hline
1.00 & $1.7 \times 10^{-9}$ & $9.0 \times 10^{10}$ & $2.0 \times 10^{12}$ & 2.9 & $-1.4$ & $-2.1$ \\
2.00 & $4.0 \times 10^{-9}$ & $7.0 \times 10^{10}$ & $2.0 \times 10^{12}$ & 3.1 & $-2.0$ & $-1.5$ \\
3.00 & $1.1 \times 10^{-8}$ & $5.0 \times 10^{10}$ & $3.0 \times 10^{12}$ & 3.1 & $-2.1$ & $-1.5$ \\
4.00 & $6.6 \times 10^{-8}$ & $5.0 \times 10^{10}$ & $2.0 \times 10^{12}$ & 2.9 & $-2.0$ & $-1.0$ \\
5.00 & $7.0 \times 10^{-7}$ & $6.0 \times 10^{10}$ & $2.0 \times 10^{12}$ & 2.5 & $-1.6$ & $-1.0$ \\
\hline\hline
\end{tabular}
\caption{Redshift-dependent parameters $(M_0, M_2, M_3, \mathfrak{a}, \mathfrak{b}, \mathfrak{c})$ used in the SFR–halo mass relation of Eq.~\eqref{eq:SFR_Mh}, following the fitting function of Ref.~\cite{Fonseca:2016qqw}. For redshifts $z>5$ we fix the parameters to their $z=5$ values, while at intermediate redshifts we interpolate the parameters between the tabulated values.}
\label{tab:SFR_params}
\end{table*}

In this section, we test the performance of the $\Q$ estimator on more realistic line-intensity maps, focusing on star formation lines that are relevant to existing and upcoming LIM experiments
(see Figure~\ref{fig:expts}). Previously, we demonstrated how $\Q$ behaves on Gaussian random fields with known cross-correlations. We now move toward a more astrophysically grounded setup by constructing intensity fields based on dark matter halo catalogs, using the publicly available code \texttt{LIMpy}\footnote{\url{https://github.com/Anirbancosmo/Limpy}} \cite{Roy:2023cpx, Roy:2023pei}. This enables us to test the robustness of the $\mathcal{Q}$ estimator on properties of realistic fields that could compromise the assumption of linear biasing between emission lines, such as varying minimum halo mass.

For this exploration, we use halos from the \href{https://www.tng-project.org/}{\textsc{IllustrisTNG300}}\footnote{\href{https://www.tng-project.org/data/downloads/TNG300-1/}{https://www.tng-project.org/data/downloads/TNG300-1/}} simulation---a large cosmological magneto-hydrodynamical simulation that follows the formation and evolution of galaxies and dark matter structures in a periodic box of side length $\approx 300$ Mpc \cite{Springel:2017tpz,Pillepich:2017fcc}. The \textsc{TNG300} run provides a statistically representative volume with sufficient halo resolution, $\approx10^{9}{\rm M}_{\odot}$, to robustly capture the halo population relevant for our analysis. We use these haloes at various redshifts for the subsequent line modeling.

The \texttt{LIMpy} code assigns line luminosities to halos through a two-step process: (1) it computes the star formation rate (SFR) from halo mass $M_h$ using the fitting formula from Ref.~\cite{Fonseca:2016qqw},
\begin{equation}
\mathrm{SFR}(M_h, z) = M_0 \left( \frac{M_h}{M_1} \right)^\mathfrak{a} \left( 1 + \frac{M_h}{M_2} \right)^\mathfrak{b} \left( 1 + \frac{M_h}{M_3} \right)^\mathfrak{c} \,,
\label{eq:SFR_Mh}
\end{equation}
where $M_1 \equiv 10^8\,{\rm M}_\odot$, and $M_0, M_2, M_3, \mathfrak{a}, \mathfrak{b}, \mathfrak{c}$ are redshift-dependent parameters specified in Table~\ref{tab:SFR_params}
(also in Table I of Ref.\cite{Fonseca:2016qqw}); and (2) it converts the resulting SFR to line luminosity $L$ via the relation \cite{Silva:2014ira}
\begin{equation}
\frac{L}{L_\odot} = 10^{\alpha_{\rm SFR}} \left( \frac{\mathrm{SFR}}{{\rm M}\odot\,{\rm yr}^{-1}} \right)^{\beta_{\rm SFR}}\,.
\label{eq:luminosity_sfr}
\end{equation}
where $L_\odot$ is the solar luminosity and the phenomenological parameters $\alpha_{\rm SFR}$ and $\beta_{\rm SFR}$ differ across lines. In addition, we impose a minimum halo mass threshold $M_{\rm min}$, below which halos are assumed incapable of hosting significant star formation and hence do not contribute to the line signal. Intuitively, $\alpha_{\rm SFR}$ sets the overall amplitude of the luminosity without affecting halo-to-halo variations across different lines, while $\beta_{\rm SFR}$ and $M_{\rm min}$ modulate the sensitivity of luminosity to SFR, effectively altering the weighting of haloes in the line intensity maps.

To understand how astrophysical modeling choices affect the performance of the $\mathcal{Q}$ estimator, we explore the parameter space spanned by $\alpha_{\rm SFR}$, 
$\beta_{\rm SFR}$, and $M_{\rm min}$. These parameters play a crucial role in shaping the statistical properties of the resulting intensity fields and can influence the cross-correlation structure that $\mathcal{Q}$ relies on. For the next section, we use a dark matter halo catalogue at $z=2$\,.

\subsection{Parameter Space Exploration}
\label{sec:param_explore}
\begin{figure*}
    \centering
    \includegraphics[width=1\textwidth]{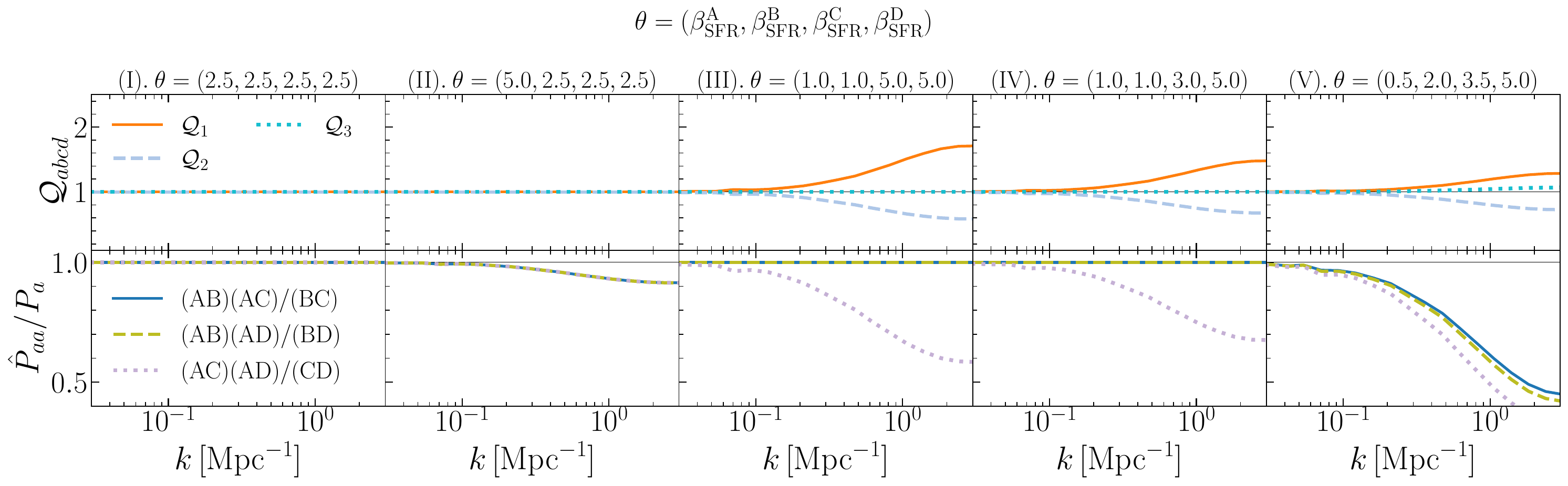}
    \caption{The $\mathbf{\mathcal{Q}}$ estimator (top row) and the \B19 estimator (bottom row) both as a function of $k$ for one of the five illustrative configurations (Cases~I–V, as described in Section~\ref{sec:param_explore}) obtained by varying the SFR–luminosity slope parameters $\beta_{\rm SFR}$ of four mock lines while keeping $\alpha_{\rm SFR}$ and $M_{\rm min}$ fixed. 
One sees that different patterns of halo weighting across the four tracers can lead to qualitatively different behaviours of the \B19\ estimator, and that no single configuration of lines is universally optimal for computing $\mathcal{Q}$: some arrangements are intrinsically more prone to false positive or false negative outcomes than others. 
}
    \label{fig:param_TF}
\end{figure*}

To assess the effectiveness of the \B19 estimator in recovering the power spectrum of line~`A', we investigate a range of physically motivated scenarios by varying the three main parameters associated with different spectral lines. Our goal is to determine under which configurations the $\mathcal{Q}(k)$ statistic reliably indicates whether the \B19 estimator succeeds or fails. We define the estimator to be successful when $\hat{P}_{aa}/P_a \approx 1$, and consider four possible outcomes of our tests. An outcome is negative if $
\mathcal{Q}(k) \approx 1$ and positive otherwise. A further classification of ``true" or ``false" is determined by whether this outcome correctly predicts the success (or not) of the \B19 estimator. These diagnostic categories are summarized in Table~\ref{tab:classification}. 
\begin{table*}[ht!]
\centering
\begin{tabular}{cccl}
\hline \hline
\textbf{Category} & $\quad \mathcal{Q}(k) \approx 1 \quad$ & $ \quad \hat{P}_{aa}/P_{a} \approx 1 \quad$ & \textbf{Interpretation} \\
\hline
{True Positive (TP)} & No & No &  $\mathcal{Q}(k)$ correctly signals the failure of the \B19 estimator.\\
{True Negative (TN)} & Yes & Yes & $\mathcal{Q}(k)$ correctly signals the success of the \B19 estimator.\\
{False Positive (FP)} & No & Yes & $\mathcal{Q}(k)$ falsely signals failure, although the \B19 estimator suceeds.\\
{False Negative (FN)} & Yes & No & $\mathcal{Q}(k)$ falsely signals success, although the \B19 estimator fails.\\
\hline \hline
\end{tabular}
\caption{Classification of the estimator performance based on the behavior of $\mathcal{Q}(k)$ and $\hat{P}_{aa}/P_{a}$.}
\label{tab:classification}
\end{table*}

We first explore five representative cases by varying $\beta_{\rm SFR}$, while keeping $\alpha_{\rm SFR}$ and $M_{\rm min}$ fixed, shown in Figure~\ref{fig:param_TF}, in order to isolate the effects of halo weighting. These configurations are
\begin{itemize}
\item \textbf{Case I}: \textbf{All lines similar,} where $\beta_{\rm SFR}^A = \beta_{\rm SFR}^B = \beta_{\rm SFR}^C = \beta_{\rm SFR}^D = 2.5$.
\item \textbf{Case II}: \textbf{Three lines similar, one different, } where $\beta_{\rm SFR}^A = 5.0, \beta_{\rm SFR}^B = \beta_{\rm SFR}^C = \beta_{\rm SFR}^D = 2.5$ and all permutations.
\item \textbf{Case III}: \textbf{Two pairs with high intra-group similarity,} where $\beta_{\rm SFR}^A = \beta_{\rm SFR}^B =  1.0$, $\beta_{\rm SFR}^C = \beta_{\rm SFR}^D= 5.0$.
\item \textbf{Case IV}: \textbf{Two similar, two distinct,} where $\beta_{\rm SFR}^A = \beta_{\rm SFR}^B= 1.0$, $\beta_{\rm SFR}^C = 3.0$, $\beta_{\rm SFR}^D = 5.0$, and \\
\item \textbf{Case-V}: \textbf{All lines different,} where $\beta_{\rm SFR}^A = 0.5, \beta_{\rm SFR}^B = 2.0, \beta_{\rm SFR}^C = 3.5, \beta_{\rm SFR}^D = 5.0$.
\end{itemize}
The results are shown in Figure~\ref{fig:param_TF}, and can be interpreted as follows:
\begin{itemize}
\item In \textbf{Case I}, all fields have identical halo weighting. This leads to $\mathcal{Q}(k) \approx 1$ and $\hat{P}_{aa}/P_{a} \approx 1$ across all $k$, resulting in a consistent \textbf{True Negative (TN)} classification.
\item In \textbf{Case II}, the mismatch in one field introduces inconsistencies at smaller scales ($k \gtrsim 0.1\,{\rm Mpc}^{-1}$), where $\mathcal{Q}(k)$ remains close to unity but $\hat{P}_{aa}/P_{a}$ deviates, leading to a \textbf{False Negative (FN)} outcome at high $k$, and \textbf{TN} at small $k$. This case is an artificially pessimistic scenario.
\item \textbf{Case III} presents both \textbf{False Positive (FP)} and \textbf{FN} behavior depending on which fields are paired. When similar lines pair, the $\mathcal{Q}(k)$ deviates while $\hat{P}_{aa}/P_{a} \approx 1$ (\textbf{FP}). Conversely, when dissimilar pairs give balanced contributions, $\mathcal{Q}(k) \approx 1$ despite estimator failure (\textbf{FN}). This case highlights the importance of considering all combinations of $\mathcal{Q}$. All of the combinations together give a good indicator of the $k$ range that the B19 estimator can be trusted but within all of the combinations there are \textbf{FN}s and \textbf{FP}s.

\item \textbf{Case IV} exhibits mixed pairings: $A$ and $B$ are similar, but $C$ and $D$ are distinct. This again leads to 
\textbf{FP} when similar lines pair, and \textbf{FN} when combinations balance out. Similar to \textbf{Case III}, all combinations of $\mathcal{Q}$ must be considered.
\item \textbf{Case-V} where all the lines are different, shows \textbf{TN} at large scales and \textbf{TP} at small scales.
\end{itemize}

We have also confirmed that varying $\alpha_{\rm SFR}$ independently, while keeping $\beta_{\rm SFR}$ and $M_{\rm min}$ fixed, always results in TPs. This is expected, since $\alpha_{\rm SFR}$ alters the overall amplitude of the luminosity without changing the relative halo contributions between lines.

Finally, we have verified that changing $M_{\rm min}$ produces similar results as changing $\beta_{\rm SFR}$, provided that $\alpha_{\rm SFR}$ and $\beta_{\rm SFR}$ are fixed across all lines. This confirms that 
$\beta_{\rm SFR}$ and $M_{\rm min}$ are the dominant parameters influencing estimator performance due to their role in modifying halo weighting, and hence the correlation between lines, which is central to the validity of the \B19 framework.

\subsection{With star-formation–tracing emission lines}

The previous section pushed the extremes of parameter space to show that TPs, TNs, FPs, and FNs are all mathematically possible. We now consider whether these possibilities happen in practice. For the remainder of this section, we adopt fiducial values of $\alpha_{\rm SFR}$ and $\beta_{\rm SFR}$ that are motivated by recent literature~\cite{Silva:2014ira, DeLooze:2014dta, Veraldi2025} and are listed in Table~\ref{tab:a_b_vals}, while fixing the minimum halo mass to $M_{\rm min} = 10^{9}\,M_\odot$. We focus on four key star-formation–tracing emission lines---\CII, \NII, \CI, and \OIII~---along with the 21-cm line in a later section, across redshifts $z < 6$, known as the post-reionization epoch. This epoch offers a unique window into the interplay between neutral hydrogen and the star-forming interstellar medium. At these redshifts, the intergalactic medium is highly ionized, with most \HI\ confined to dense, self-shielded regions within galaxies. Several ongoing and upcoming intensity mapping experiments target this era: CHIME\footnote{\href{https://chime-experiment.ca/}{https://chime-experiment.ca/}},
CHORD\footnote{\href{https://www.chord-observatory.ca/}{https://www.chord-observatory.ca/}},
and SKA-Mid\footnote{\href{https://www.skao.int/en/explore/telescopes/ska-mid}{https://www.skao.int/en/explore/telescopes/ska-mid}} aim to detect the 21-cm signal; 
COMAP\footnote{\href{https://comap.caltech.edu/}{https://comap.caltech.edu/}}
and COPSS \cite{Keating:2015qva} focus on CO rotational lines; 
and FYST\footnote{\href{http://www.ccatobservatory.org/}{http://www.ccatobservatory.org/}}
and TIME\footnote{\href{https://sites.google.com/view/abigailtcrites/research/time}{https://sites.google.com/view/abigailtcrites/research/time}}
are expected to measure far-infrared lines such as \CII, \CI, and \NII. 
The overlap of the observational windows of these observations in the post-reionization era
makes it an ideal regime for joint analyses and cross-correlation studies involving multiple tracers of large-scale structure (see Figure~\ref{fig:expts}).

In this context, we assess the feasibility of using the $\mathcal{Q}$ estimator to evaluate the applicability of the \B19 estimator based on these four lines. We consider two observational scenarios: one using idealized line intensity maps, and another incorporating realistic instrumental noise characteristics.

\begin{table}[ht]
\centering
\begin{tabular}{lcc}
\hline\hline
\textbf{Line} & \boldmath\( \alpha_{\rm SFR} \) & \boldmath\( \beta_{\rm SFR} \) \\
\hline
{\CII} $-$ 158\,µm   & 6.98 & 0.99 \\
{\CI} $-$ 371\,µm    & 6.30 & 0.50 \\
{\OIII} $-$ 88\,µm   & 7.40 & 0.97 \\
{\NII} $-$ 205\,µm   & 5.70 & 0.95 \\
\hline\hline
\end{tabular}
\caption{Line-luminosity relation parameters \( \alpha_{\rm SFR} \) and \( \beta_{\rm SFR} \) used for generating intensity maps (Eq.~\eqref{eq:luminosity_sfr}). Here the numbers after the molecular symbols represent the
wave-lengths of the corresponding lines in microns. In the text, for brevity, we 
represent a line with its molecular symbol and this wavelength number only}
\label{tab:a_b_vals}
\end{table}

\subsubsection{Pure Signal}

\begin{figure*}
    \centering
    \includegraphics[width=1\linewidth]{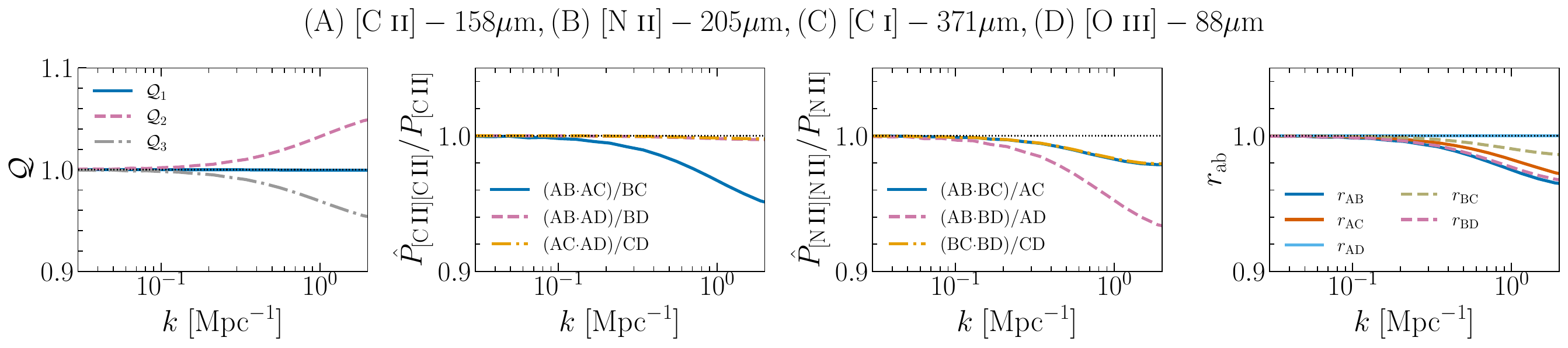}
    \caption{Performance of the $\Q$–estimator and the \B19\ power–spectrum estimator for four star–formation–tracing lines (\CII, \NII, \CI, and \OIII) in the absence of instrumental noise at $z=2$. 
The left panel shows the three $\Q$ combinations, $\Q_1$, $\Q_2$, and $\Q_3$ (defined in Eq.~\eqref{eq:Qi_defs}), as functions of $k$. 
On large scales all three combinations are consistent with $\Q\simeq1$, while $\Q_2$ and $\Q_3$ begin to deviate at the few percent level for $k\gtrsim0.2\,\mpci$, reflecting small differences in halo weighting between the tracers. 
The middle and right panels show the ratios of the \B19–reconstructed power spectra to the true spectra for \CII\ and \NII, respectively, for all tri–line combinations: the \B19\ estimator is accurate to better than $\sim5\%$ on scales where the corresponding $\Q_i$ remain close to unity, and gradually departs from unity once the $\Q$ combinations begin to drift. 
The rightmost panel displays the cross–correlation coefficients $r_{ab}(k)$ between all pairs of lines, demonstrating that the tracers are very highly correlated ($r_{ab}\simeq1$) at $k\lesssim0.2\,\mpci$, with correlation degrading on smaller scales where the departures in both $\Q$ and \B19\ become visible.
}
    \label{fig:LIM_no_noise}
\end{figure*}

In this section, we focus exclusively on star formation–tracing emission lines (\CII, \NII, \CI, and \OIII) to test the performance of the $\mathcal{Q}$-estimator.
Figure~\ref{fig:LIM_no_noise} presents the performance of the $\mathcal{Q}$-estimator for the four selected lines using idealized, noise-free line-intensity maps at $z = 2$. 
We generate the intensity maps of these lines using
\texttt{LIMpy}.
The panels show the scale dependence of $\mathcal{Q}$, the recovered power spectra using \B19 estimator for \CII\ and \NII, and cross-correlation coefficients among the different line combinations. Here we chose three different combinations of the cross power spectrum to compute the $\Q$ estimator, namely $\Q_1,\Q_2$, and $\Q_3$ and similarly for the \B19 estimator outlined in Section \ref{sec:formalism}. The combinations are labelled in the figures. The cross-correlation coefficients between lines are computed as
\begin{equation}
    r_{ab} = \frac{P_{ab}}{\sqrt{ P_{aa} P_{bb}}}\,.
\end{equation}
As discussed earlier, for ideal, perfectly correlated, and linearly biased fields, we expect $\mathcal{Q} = 1$. In Figure~\ref{fig:LIM_no_noise} we find that $\mathcal{Q}_1$ remains close to unity across the entire $k$ range, while $\mathcal{Q}_2$ and $\mathcal{Q}_3$ begin to deviate from unity beyond $k > 0.2\,\mpci$, with discrepancies of about 5\%. Examining the \B19 estimated power spectra for \CII\ (second panel) and \NII\ (third panel), we observe deviations from the true spectra at $k > 0.2\,\mpci$ by less than 5\%. Although the purple and yellow curves in the second panel appear to coincide with unity, a subtle divergence emerges beyond $k > 0.2\,\mpci$.

Turning to the cross-correlation coefficients, we find that \CII\ and \OIII\ are strongly correlated, exhibiting $r_{ab} \approx 1$ throughout the $k$ range. For other line combinations, $r_{ab}$ remains close to unity for $k < 0.2\,\mpci$ but departs at higher $k$. These deviations likely stem from minor differences in halo weighting between lines, which manifest as small departures from unity in both the $\Q$ estimators and the \B19 power spectrum estimates, introduced by the weak stochasticity and scale dependence
in the bias on non-linear scale.

Figure~\ref{fig:LIM_no_noise} highlights an important point: all $\mathcal{Q}$ combinations should be considered collectively to draw reliable conclusions about the \B19 estimation. Moreover, in tri-line combinations where any two lines are highly correlated, the \B19 estimation for one of the correlated lines tends to agree more closely with the true power spectrum. Finally, at small $k$, the stability of $\mathcal{Q}$ indicates that it is a robust estimator, largely insensitive to cosmic variance.
Appendix~\ref{sec:variance} shows that in the perfectly correlated, linear-bias limit (\(r_{ab}\to 1\) for all pairs)
and without instrumental noise,\(\mathrm{Var}\,\mathcal{Q}\) vanishes:
\[
\frac{\mathrm{Var}\,\mathcal{Q}}{\mathcal{Q}^{2}}
\;\xrightarrow{\,r_{ab}\to1\,}\;0.
\]
This explains the observed stability of \(\mathcal{Q}\) on large scales, even though cosmic variance for each power spectrum is largest at small \(k\), it cancels in the
ratio because the same long-wavelength modes modulate the numerator and
denominator in a nearly identical way. Residual departures of
\(\mathcal{Q}\) from unity at high \(k\) arise when
\(r_{ab}<1\) due to non-linear and tracer-dependent halo weighting, in which case
\(\mathrm{Var}\,\mathcal{Q}/\mathcal{Q}^{2}\sim N_m^{-1}(1-r_{ab})^{2}\), where $N_m$ is the number of independent modes that go into a particular $k$ bin. This represents a small but
non-zero departure that grows as correlation degrades.

\begin{figure*}
    \centering
    \includegraphics[width=1\linewidth]{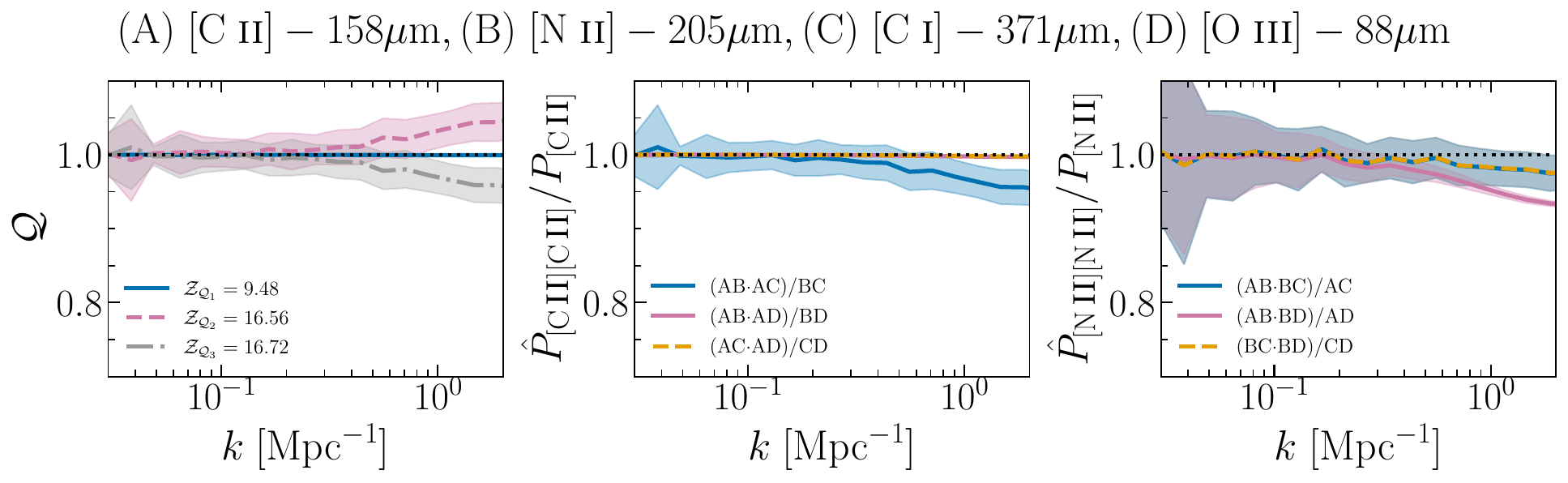}
    \caption{Same as Fig.~\ref{fig:LIM_no_noise}, but including Gaussian instrumental noise appropriate for the fiducial Super–LIM configuration with $N_{\rm det}\,t_{\rm survey}=2\times10^5\,{\rm hr}$. 
The left panel shows $\Q_1$, $\Q_2$, and $\Q_3$ as functions of $k$, with shaded bands indicating the $1\sigma$ scatter over many noise realizations. 
On large scales ($k\lesssim0.3\,\mpci$) all three combinations remain consistent with $\Q=1$ within the uncertainties, while at smaller scales $\Q_2$ and $\Q_3$ exhibit statistically significant departures, signalling the breakdown of the assumptions underlying the \B19\ estimator. 
The middle and right panels show the \B19–reconstructed power spectra for \CII\ and \NII\ (normalized by the true spectra) for different tri–line combinations; for \CII\ the reconstruction remains accurate up to $k\simeq0.3\,\mpci$, whereas for the fainter \NII\ line the deviations become more pronounced and combination–dependent. 
The scales where $\Q_2$ and $\Q_3$ depart from unity coincide with the onset of bias in the \B19\ reconstructions, illustrating the utility of $\Q$ as a noise–robust null test.
}
    \label{fig:LIM_noise}
\end{figure*}

\subsubsection{Including Instrument Noise}
We next assess the robustness of the \(\Q\) estimator in the presence of realistic instrumental noise. To this end, we consider a hypothetical single-dish experiment, Super-LIM, capable of simultaneously observing multiple far-infrared (FIR) emission lines, including \CII, \NII, \CI, and \OIII, and having a frequency coverage of $10$ to $1000$ GHz. The instrumental setup for Super-LIM is designed to capture the key characteristics of upcoming LIM experiments such as FYST, CONCERTO, EXCLAIM, and TIME. The details of the assumed instrument configuration and the methodology for generating noisy realizations of the intensity maps are described in Appendix~\ref{sec:fir_noise}. 
The root-mean-square (RMS) noise amplitude, $\sigma_{\rm rms}$ (Equation~\ref{eq:rms-lim}), scales inversely with $(N_{\rm det}, t_{\rm survey})^{1/2}$, where $N_{\rm det}$ is the number of detectors and $t_{\rm survey}$ is the total survey duration in hours. Throughout the text, we report the combined parameter $N_{\rm det} t_{\rm survey}$ when specifying the noise level for Super-LIM.

We generate 100 independent noisy realizations of the intensity maps by first creating 100 Gaussian noise maps for the Super-LIM instrument and adding them to the corresponding simulated line intensity maps. For each realization, we compute all relevant quantities, including the $\mathcal{Q}$ estimator. 
 These realizations are then used to estimate the expected mean and variance.

Figure~\ref{fig:LIM_noise} shows the resulting measurements based on these noise-augmented maps.
For this we choose $N_{\rm det} t_{\rm survey}=2\times10^5$ hrs.  
From the left panel, we see that $\mathcal{Q}$ remains close to unity at large scales ($k \lesssim 0.3\mpci$), confirming the consistency of the estimator under realistic noise conditions. However, increased fluctuations are visible toward both ends of the $k$-range. At low $k$ ($k \lesssim 0.02\mpci$), the limited number of modes ($N_m(k)\propto k^{2}$) amplifies the impact of residual noise, resulting in larger variance. 
However, as seen in Figure~\ref{fig:LIM_noise}, $\mathcal{Q}_2$ and $\mathcal{Q}_3$ exhibit statistically significant deviations from unity at small scales ($k > 0.5\mpci$). This demonstrates the sensitivity of the $\mathcal{Q}$ estimator as an effective null test for identifying the departure from the \B19 assumptions even in the presence of realistic noise. To quantify this departure more rigorously, we define a statistic, $\z_{\Q}$,
\begin{equation}
\z_{\Q} = \sum_{k_i} \left[\frac{\mathcal{Q}(k_i)-1}{\Delta\mathcal{Q}(k_i)}\right]^2,
\label{eq:chi_def}
\end{equation}
where the sum runs over the $k$ bins, and $[\Delta\mathcal{Q}(k_i)]^2$ represents the variance in each bin estimated from independent noise realizations. This statistic quantifies the overall significance of deviations from unity, providing a direct measure of the power of $\mathcal{Q}$ as a diagnostic test. The square root of this quantity,  $\sqrt{\z_{\Q}}$, roughly determines the significance (``number of sigmas") of the departure from unity. We find that $\Q_2$ and $\Q_3$ capture the deviations from unity with a significance of approximately $4\,\sigma$, indicating strong sensitivity to departures from the \B19\ assumptions. In contrast, $\Q_1$ shows a lower significance of about $3\,\sigma$, remaining close to unity with notably smaller variance than the other two. This behavior suggests that $\Q_1$ is less effective as an indicator of deviation, reinforcing the importance of considering all $\Q$ combinations together when assessing consistency. 

In the middle and right panels, the \B19\ estimator continues to reproduce the power spectra accurately up to $k < 0.3\,\mpci$, where the signal-to-noise ratio (SNR) remains high. The performance is particularly strong for the brighter lines, such as \CII, and for combinations involving multiple bright tracers. In contrast, the fainter \NII\ line shows lower significance, consistent with its reduced SNR. The departure from the true power spectrum becomes evident for \NII\ across all \B19\ combinations, while for \CII\ this is noticeable in only one case. Notably, the $k$ values where these deviations appear coincide with the scales at which $\mathcal{Q}$ begins to depart significantly from unity. 
This reinforces the effectiveness of $\mathcal{Q}$ as a diagnostic quantity even under realistic instrumental noise conditions.

\begin{figure*}
    \centering
    \includegraphics[width=1\textwidth]{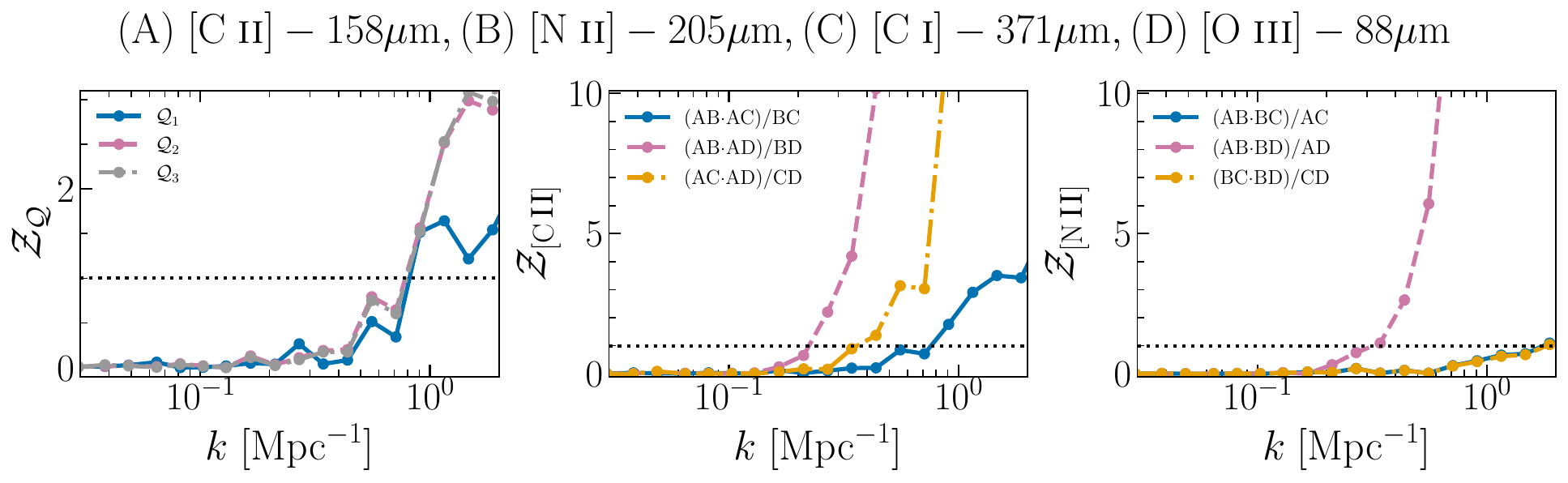}
    \caption{Scale-dependent significance $\mathcal{Z}_X(k)$ of departures from the ideal relations for the noisy four–line LIM configuration at $z=2$ with $N_{\rm det}\,t_{\rm survey}=2\times10^5\,{\rm hr}$. 
Left: $\mathcal{Z}_{\Q}(k)$ for $\Q_1$, $\Q_2$, and $\Q_3$; all combinations are consistent with $\Q=1$ at large scales, while $\Q_2$ and $\Q_3$ exceed the $1\sigma$ threshold (horizontal dotted line) on small scales, reaching several–$\sigma$ significance in the strongly non–linear regime. 
Middle: $\mathcal{Z}(k)$ for the \B19\ reconstruction of the \CII\ power spectrum for all tri–line combinations; the most informative combinations show rapidly growing tension with the true spectrum once $k\gtrsim0.3\,\mpci$, while others remain closer to unity. 
Right: analogous statistic for \NII, where combinations involving bright tracers yield strong detections of inconsistency, whereas other choices of lines remain only marginally significant. 
The close correspondence between the scales at which $\mathcal{Z}_{\Q}$ and the \B19\ statistics exceed $1\sigma$ demonstrates that the $\Q$–estimator provides a sensitive per–mode null test for the validity of the \B19\ assumptions.
}
    \label{fig:Z_by_k_1000}
\end{figure*}

Figure~\ref{fig:Z_by_k_1000} examines the scale-by-scale sensitivity of the $\Q$-estimator and of the \B19\ power-spectrum estimator. For each $k$-bin we define a local statistic
\begin{equation}
\mathcal{Z}_{X}(k_i) \equiv
\left[\frac{X(k_i)-X_{\rm fid}(k_i)}{\Delta X(k_i)}\right]^2\,,
\label{eq:Zk}
\end{equation} 
where $X$ denotes either one of the $\Q_i$ combinations or a \B19-recovered power-spectrum. 
The fiducial value $X_{\rm fid}(k)$ represents the expected result when the assumptions of the corresponding estimator hold: for the $\Q$-statistics we have $X_{\rm fid}=1$, while for the \B19\ estimator $X_{\rm fid}(k)$ is the true underlying auto-power spectrum $P_{a}(k)$.  
The quantity $\Delta X(k_i)$ is the standard deviation across the noisy realizations.
The horizontal dotted line in each panel marks $\mathcal{Z}_X=1$, corresponding to a $1\sigma$ local deviation.

The left panel shows $\mathcal{Z}_{\Q}(k)$ for the three combinations $\Q_1$, $\Q_2$, and $\Q_3$ defined in Eq.~\eqref{eq:Qi_defs}. On large scales ($k\lesssim0.3\,\mpci$) all three curves lie well below the $1\sigma$ threshold, indicating that the measured $\Q_i$ are fully consistent with unity once noise is accounted for. At intermediate scales, $\Q_2$ and $\Q_3$ begin to rise, crossing the $1\sigma$ line around $k\simeq0.7$ to $0.8\,\mpci$ and reaching $\mathcal{Z}_{\Q}\sim3$ at the smallest scales probed. In contrast, $\Q_1$ remains below this threshold over most of the range and only approaches the $1\sigma$ level at the largest $k$. This confirms the earlier qualitative impression that $\Q_1$ is comparatively conservative, while $\Q_2$ and $\Q_3$ are more responsive to departures from the \B19\ assumptions on non-linear scales.

The middle and right panels show the corresponding $\mathcal{Z}$-statistics for the \B19\ power-spectrum estimates of \CII\ and \NII, respectively. For \CII\ (middle panel), each curve represents one tri-line combination used in the estimator. 
All combinations yield $\mathcal{Z}_{\textrm{C}\textsc{ii}}(k)<1$ in the range $k\sim0.3$ to $0.6\,\mpci$, confirming unbiased recovery of the \CII\ power spectrum there. 
Beyond this range, however, some combinations exhibit very rapid growth in $\mathcal{Z}_{\textrm{C}\textsc{ii}}$: the most sensitive ones exceed the $3\sigma$ ($\z$ value $\gtrsim10$) level by $k\sim0.5\,\mpci$ and goes beyond the plotting range ($\gtrsim10\sigma$) at still smaller scales. The remaining combination rises more slowly and only reaches a few-sigma tension at the highest $k$. This spread reflects the different ways in which each tri-line combination weights the bright and faint tracers, and shows that some choices of lines provide a much more stringent measurement of the auto-power spectrum.

For \NII\ (right panel), the behaviour is even more heterogeneous. The combination involving (\CII, \NII, \OIII) yields a very strong detection of inconsistency, with $\mathcal{Z}_{\textrm{N}\textsc{ii}}$ jumping above $1\sigma$ around $k\simeq0.4\,\mpci$ and rapidly exceeding the $3\sigma$ range at higher $k$. In contrast, the other two combinations remain consistent with unity to within $1\sigma$ over most of the $k$-range and only reach marginal ($\sim1\sigma$) deviations at the smallest scales. This pattern mirrors the lower signal-to-noise of the \NII\ line: only those combinations where \NII\ appears with two other strong lines achieve high sensitivity to deviations from the \B19\ relations.

Taken together, the three panels demonstrate that the scales at which $\mathcal{Z}_{\Q}(k)$ first exceeds unity closely track the scales where at least one \B19\ combination for \CII\ or \NII\ shows a significant departure from the true power spectrum. The $\Q$-estimator therefore functions as a robust null test: as long as $|\mathcal{Z}_{\Q}(k)|\lesssim1$ for all combinations, the \B19\ reconstruction remains statistically consistent with the underlying spectra; once $\Q_2$ and $\Q_3$ cross the $1\sigma$ threshold, several \B19\ combinations simultaneously enter the multi-$\sigma$ regime. At the same time, the differing behaviour of $\Q_1$, $\Q_2$, and $\Q_3$ underscores that no single combination provides a complete diagnostic: only by considering all of them, in parallel with the various \B19\ line combinations, can we fully characterise the breakdown of the underlying assumptions and identify the most informative sets of tracers.

\begin{figure*}
    \centering
    \includegraphics[width=1\textwidth]{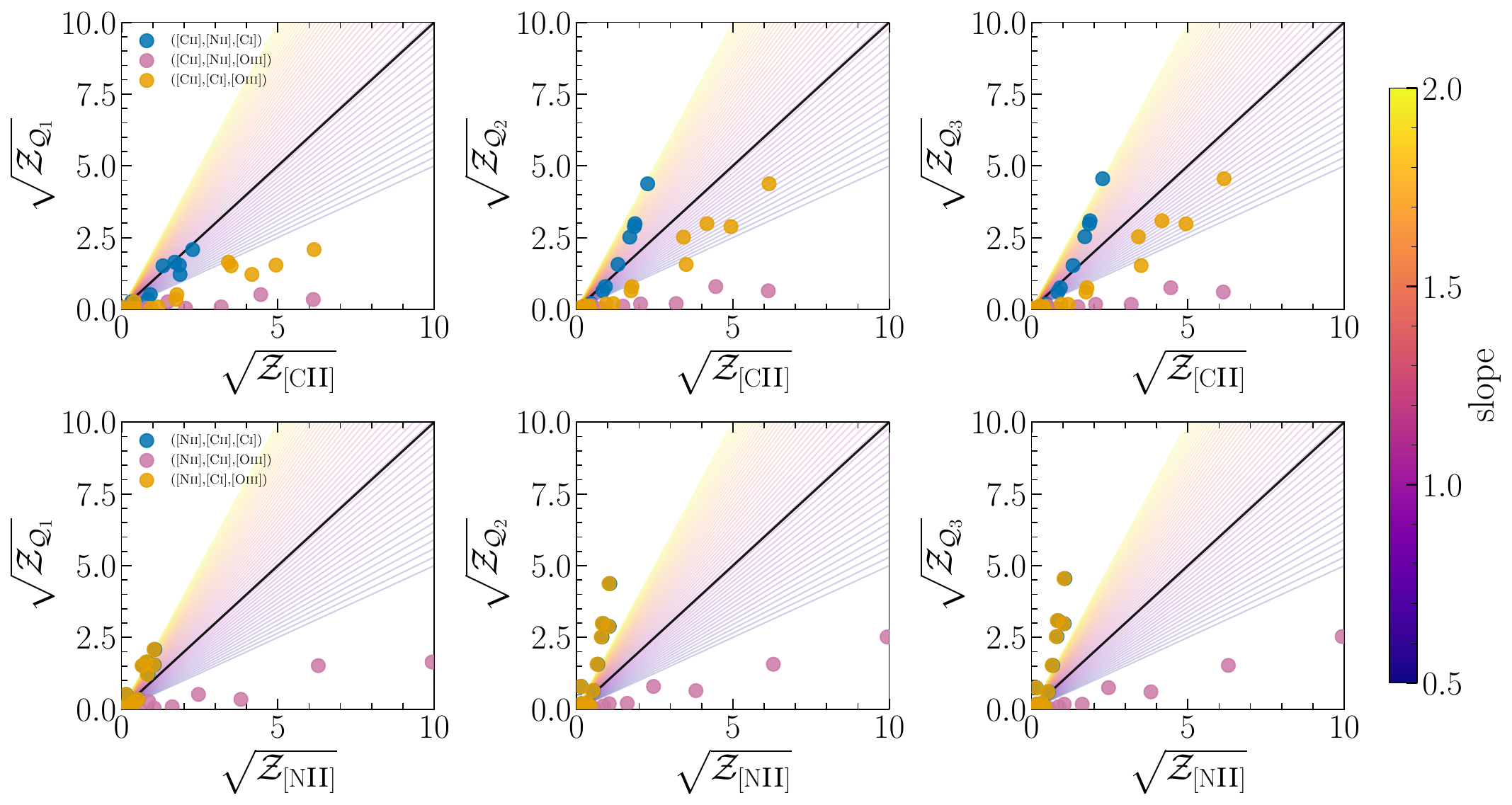}
    \caption{Global comparison of the detection significance from the $\Q$–estimator and from the \B19\ power–spectrum estimator for the noisy four–line LIM configuration at $z=2$ and $N_{\rm det}\,t_{\rm survey}=2\times10^5\,{\rm hr}$. 
Each panel shows $\sqrt{\z_{\Q_i}}$ versus $\sqrt{\z_{\rm line}}$, where $\z_X$ is the $k$–summed statistic defined in Eq.~\eqref{eq:chi_def}, the horizontal axis corresponds to either \CII\ (top row) or \NII\ (bottom row), and the vertical axis to one of the three $\Q_i$ combinations (columns from left to right). 
Points represent individual tri–line choices entering the \B19\ estimator, colour–coded by which set of lines is used, while the radial coloured fans mark lines of constant slope $m=\sqrt{\z_{\Q_i}}/\sqrt{\z_{\rm line}}$. 
For \CII, most configurations lie near slopes of order unity, indicating that the global significance of the deviations inferred from $\Q_2$ and $\Q_3$ closely tracks that of the \B19\ estimator, and can even exceed it for favourable line combinations. 
For \NII, the spread in slopes is larger because of the lower SNR of this line, but the most informative triads still yield $\Q$–based significances comparable to those from \B19. 
Overall, the figure highlights that $\Q_1$, $\Q_2$, and $\Q_3$ provide complementary diagnostics that correlate strongly with the \B19\ power–spectrum deviations, and should be used jointly to assess internal consistency.
}
    \label{fig:scatter_LIM_1000}
\end{figure*}

Figure~\ref{fig:scatter_LIM_1000} presents an alternative view of the same information shown in Figure~\ref{fig:Z_by_k_1000}. The horizontal axes show $\sqrt{\z_{[\rm C \textsc{ii}]}}$ (top row) or $\sqrt{\z_{[\rm N \textsc{ii}]}}$ (bottom row), i.e.\ the significance of the deviation of the \B19\ power-spectrum estimate for \CII\ or \NII\ from the true spectrum. The vertical axes show $\sqrt{\z_{\Q_i}}$ for the three $\Q$ combinations ($i=1,2,3$; columns from left to right). Each point corresponds to one choice of tri-line combination, with colours indicating which set of lines enters the estimator as listed in the legends.

The coloured fans of straight lines emanating from the origin indicate lines of constant slope, with the colour bar encoding the value of the slope. A point lying on a line with slope $m$ satisfies
\begin{equation}
    \sqrt{\z_{\Q}} = m \sqrt{\z_{{\rm line}}}
\end{equation}
so that $m$ directly measures the relative sensitivity of $\Q_i$ compared to the \B19\ estimator for that configuration: $m>1$ (red) implies that $\Q_i$ detects the breakdown of the \B19\ assumptions more significantly than the direct comparison of the recovered and true power spectra, while $m<1$ (blue) indicates that the \B19\ deviation is more significant. Essentially, $m$ is a more quantitative, nuanced way to distinguish between FPs and FNs.

In the top row, where the horizontal axis corresponds to \CII, many points cluster around slopes of order unity for all three $\Q_i$, indicating that the significance of the deviations inferred from $\Q$ tracks that from the \B19\ \CII\ spectrum. For several configurations, particularly those involving a bright tracer combination (blue and yellow points), the inferred slopes are $m\gtrsim1$, showing that $\Q_2$ and $\Q_3$ in particular can be more sensitive to departures from the \B19\ relations than the direct power-spectrum comparison. Only a minority of points fall well below the $m=0.5$ line, reflecting cases where the \B19\ estimator for \CII\ becomes strongly inconsistent while the corresponding $\Q_i$ combination still yields only modest significance.

The bottom row, which compares $\Q_i$ against the \B19\ estimator for \NII, exhibits a broader spread. Several pink points lie far along the horizontal axis with only moderate vertical displacement, corresponding to configurations where the \NII\ power spectrum reconstructed by \B19\ is in strong ($\gtrsim5\sigma$) tension with the truth, while $\Q_i$ registers a weaker ($\lesssim2$–3$\sigma$) deviation.  A subset of configurations still lie near or above the $m=1$ line, demonstrating that when \NII\ participates in sufficiently strong cross-correlations, the $\Q$-estimator remains a competitive or superior diagnostic.

Overall, this figure confirms the conclusions drawn from the $k$-resolved $\mathcal{Z}$-statistics: (i) for bright tracers such as \CII, the $\Q$-estimator and the \B19\ power-spectrum comparison yield comparable global significances, with $\Q_2$ and $\Q_3$ often providing an equally or more sensitive null test; (ii) for noisier lines such as \NII, the sensitivity of $\Q$ becomes more configuration-dependent, but still tracks the \B19\ deviations for the most informative tri-line combinations. Thus, the scatter plots emphasise that $\Q_1$, $\Q_2$, and $\Q_3$ should be used jointly with multiple \B19\ combinations: taken together, they map out how strongly the internal consistency of the multi-line data set is violated and identify which tracer combinations provide the most powerful diagnostics.

\begin{figure*}
    \centering
    \includegraphics[width=1\textwidth]{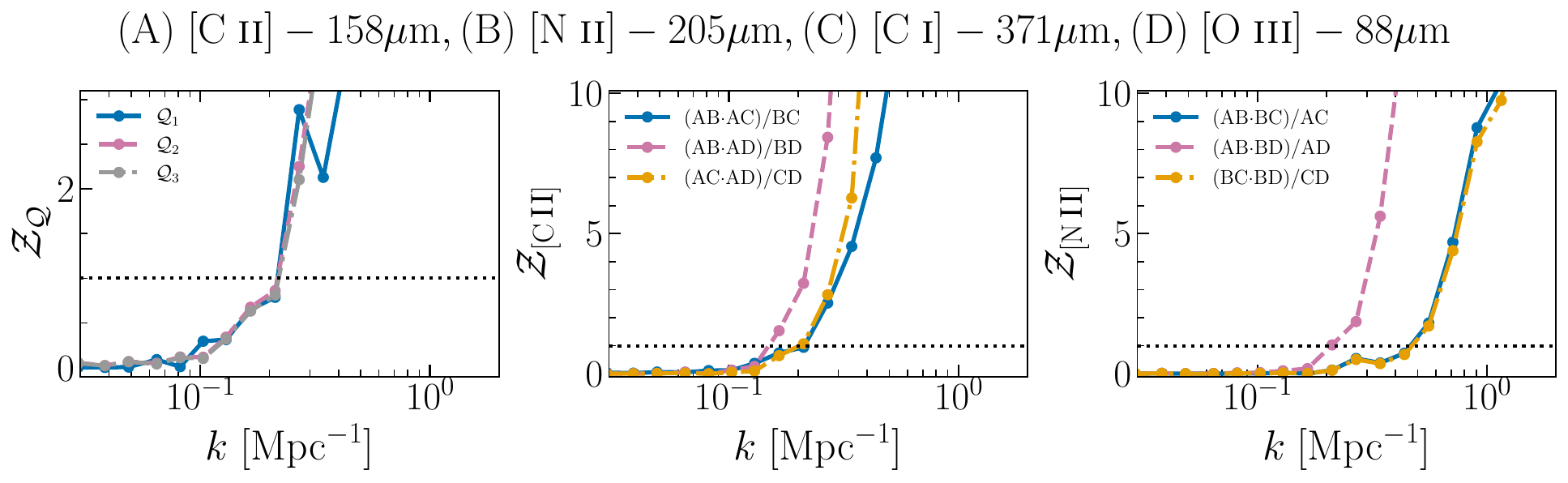}
    \caption{Same as Fig.~\ref{fig:Z_by_k_1000}, but for a deeper Super–LIM configuration with a larger effective exposure ($N_{\rm det}\,t_{\rm survey}=10^6\,{\rm hr}$ and correspondingly lower noise). 
The left panel shows that the per–bin statistics $\mathcal{Z}_{\Q_i}(k)$ change only mildly compared to the shallower survey: the $\Q_i$ combinations still remain consistent with unity on large scales and become significant only in the mildly and strongly non–linear regimes. 
By contrast, the middle and right panels demonstrate that the per–bin significances for the \B19–reconstructed \CII\ and \NII\ power spectra grow substantially at fixed $k$, with several combinations entering the multi–$\sigma$ regime already at moderate wavenumbers. 
This behaviour reflects the fact that, as instrumental noise is reduced, the same underlying systematic departures between \B19\ and the true power spectra become statistically much more significant, while the $\Q$ combinations retain similar scale–dependence. 
The figure thus confirms that the qualitative relationship between $\Q$ and \B19\ persists as survey sensitivity improves.
}
    \label{fig:Z_by_k_5000}
\end{figure*}

Next we repeat the $\mathcal{Z}$–statistic analysis for a deeper \textsc{Super-LIM} configuration with an increased effective exposure
$(N_{\rm det} t_{\rm survey})^{1/2}=10^{6}\,{\rm hr}$ (and hence a lower
$\sigma_{\rm rms}\propto (N_{\rm det} t_{\rm survey})^{-1/2}$). The left panel of Figure~\ref{fig:Z_by_k_5000} shows that the
behaviour of the three $\Q_i$ combinations is very similar to the shallower
case in Figure~\ref{fig:Z_by_k_1000}. All three curves remain consistent with zero at large scales and cross
the $1\sigma$ threshold at nearly the same wavenumber as before. The maximum
values of $\mathcal{Z}_{\Q}(k)$ at the smallest scales increase only mildly.

By contrast, the middle and right panels, which show $\mathcal{Z}_{[\textrm{C}\,\textsc{ii}]}(k)$
and $\mathcal{Z}_{[\textrm{N}\,\textsc{ii}]}(k)$ for the different \B19\ combinations, exhibit a
much stronger response to the longer integration. For both lines, the scale at
which the curves cross the $1\sigma$ line moves to slightly larger scales, and
the absolute $\mathcal{Z}$-values at fixed $k$ increase noticeably. In particular, all
three \CII\ combinations now reach multi–$\sigma$ tension already at
moderate $k$, and even the previously conservative combinations involving
\NII\ climb into the several–$\sigma$ regime at the smallest scales. This is
exactly the expected behaviour when the dominant uncertainty in the recovered
power spectra is instrumental: as the noise decreases, the same underlying
systematic deviations between the \B19\ estimate and the true spectrum become
more prominent.

\begin{figure*}
    \centering
    \includegraphics[width=1\textwidth]{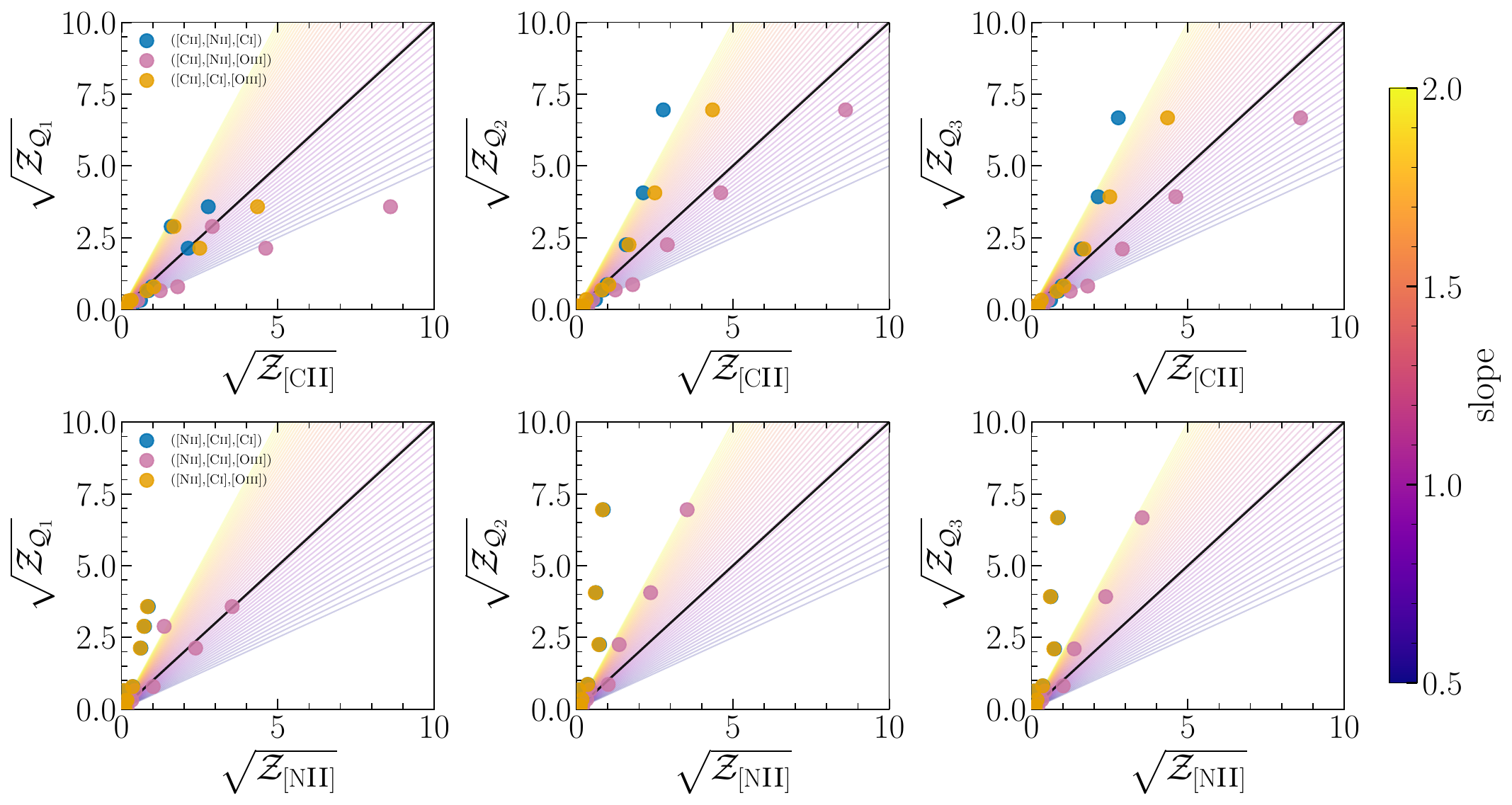}
    \caption{Global comparison of the $\Q$– and \B19–based significances for the deeper Super–LIM configuration ($N_{\rm det}\,t_{\rm survey}=10^6\,{\rm hr}$), in the same format as Fig.~\ref{fig:scatter_LIM_1000}.  
Compared to the shallower survey, the points move systematically up and to the right along nearly radial trajectories, indicating that both the $\Q$– and \B19–based significances increase together as noise is reduced, while their ratio (the slope) remains nearly unchanged. 
For \CII, most configurations continue to cluster near slopes of order unity, confirming that the $\Q$–estimator remains as sensitive as the direct \B19–truth comparison in diagnosing inconsistencies. 
For \NII, the broader spread in slopes persists, but the deepest configurations show that $\Q_2$ and $\Q_3$ can still track the \B19\ deviations for the most informative combinations. 
This figure reinforces the conclusion that the three $\Q_i$ combinations provide a stable, noise–resilient set of null tests whose global significances scale consistently with those of the \B19\ power–spectrum deviations as survey depth increases.
}
    \label{fig:scatter_LIM_5000}
\end{figure*}

The corresponding scatter plots of $\sqrt{\z_{\Q_i}}$ versus
$\sqrt{\z_{\rm line}}$ in  Figure~\ref{fig:scatter_LIM_5000} 
for the deeper survey show that the points move
systematically up and to the right, along roughly radial directions of nearly
constant slope. In other words, both the global significance of the departures
in $\Q$ and in the \B19\ power spectra increase, while their ratio is largely
preserved. For \CII, most tri-line configurations still lie close to slopes of
order unity for all three $\Q_i$, confirming that the $\Q$-estimator remains as
sensitive as the direct \B19–truth comparison in diagnosing inconsistencies,
with some configurations (especially for $\Q_2$ and $\Q_3$) yielding
$\sqrt{\z_{\Q_i}} \gtrsim \sqrt{\z_{[\rm C\textsc{ii}]}}$. For \NII, the broader
spread in slopes persists, reflecting the lower intrinsic SNR of this line,
but several points shift towards higher $\sqrt{\z_{\Q_i}}$ at fixed
$\sqrt{\z_{[\rm N\textsc{ii}]}}$, indicating that the deeper data allow $\Q$ to
better track the breakdown of the \B19\ relations for the most informative
tri-line combinations.

Overall, increasing the observing time primarily boosts the significance with
which both diagnostics detect departures from the \B19\ assumptions, rather
than qualitatively changing their relative performance. 

\begin{figure*}
    \centering
    \includegraphics[width=1\textwidth]{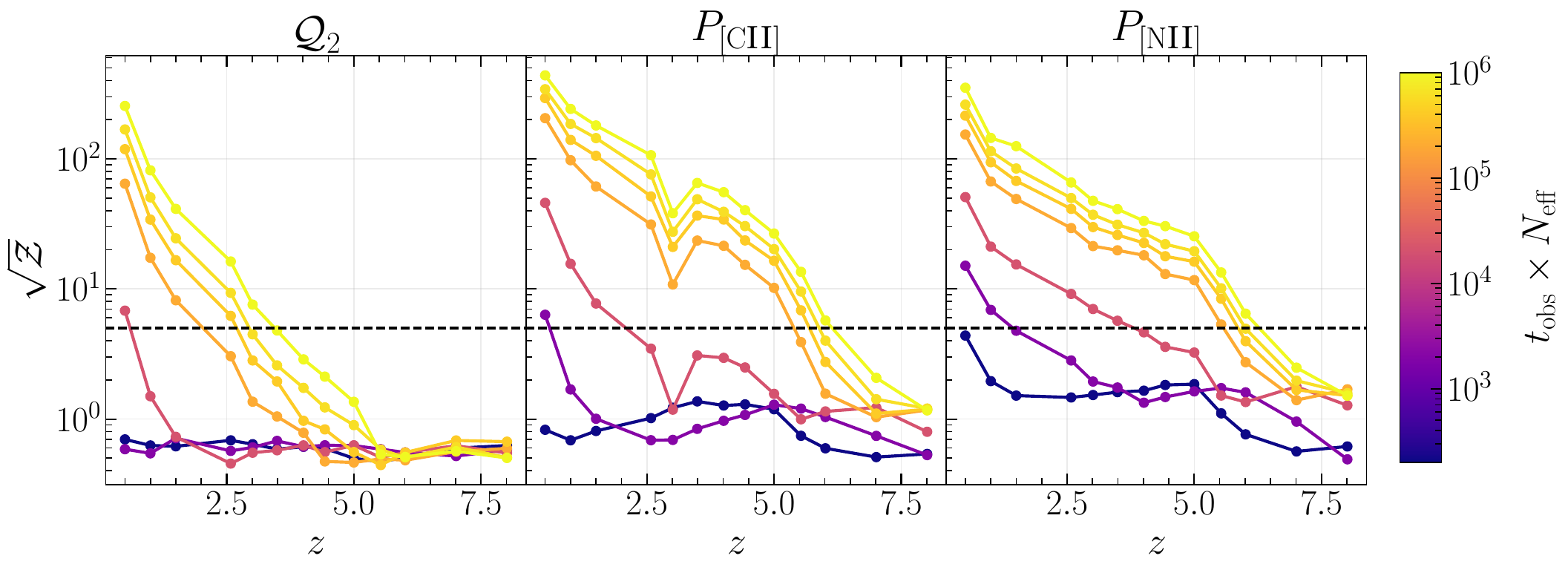}
    \caption{Redshift and survey–depth dependence of the global detection significance for a representative four–line LIM configuration, using three diagnostics. 
For each redshift we compute the $k$–summed statistic $\z_X$ of Eq.~\eqref{eq:chi_def} and plot its square root, $\sqrt{\z_X}$, for $\Q_2$ (left panel), the \B19\ reconstruction of the \CII\ power spectrum (middle), and the \B19\ reconstruction of the \NII\ power spectrum (right). 
Coloured curves correspond to different effective exposures $t_{\rm obs}N_{\rm eff}$ for the Super–LIM experiment (colour scale), which controls the instrumental noise level via $\sigma_{\rm rms}\propto(t_{\rm obs}N_{\rm eff})^{-1/2}$. 
At fixed depth, all three diagnostics show a strong redshift dependence: the significance is very high at low redshift and declines rapidly as the lines dim and noise becomes more important, with \CII\ remaining sensitive to higher $z$ than \NII. 
At fixed redshift, increasing $t_{\rm obs}N_{\rm eff}$ raises the significances of both $\Q_2$ and the \B19\ estimators, pushing them above the nominal $5\sigma$ threshold (horizontal dashed line) over an extended redshift range. 
This figure summarises the redshift window and survey depths over which multi–line LIM data can robustly diagnose breakdowns of the \B19\ assumptions.
}
    \label{fig:Z_q_time_z_LIM}
\end{figure*}

Figure~\ref{fig:Z_q_time_z_LIM} summarises how the detection significance of the departures from the \B19\ assumptions evolves with redshift and survey depth for one representative tri-line configuration. Here we take slightly larger $z$ range, $1\le z \le8$, to demonstrate the results.
For each redshift bin we compute the $\z$-statistic defined in Eq.~\eqref{eq:chi_def}, summing over all $k$-modes, and plot its square root, $\sqrt{\z_X}$, for three diagnostics: $\Q_2$ (left), the \B19\ power-spectrum estimate for \CII\ using the combination $([\textrm{C}\,\textsc{ii}],[\textrm{N}\,\textsc{ii}],[\textrm{O}\,\textsc{iii}])$ (middle), and the \B19\ estimate for \NII\ using $([\textrm{N}\,\textsc{ii}],[\textrm{C}\,\textsc{ii}],[\textrm{O}\,\textsc{iii}]$) (right). The colour scale shows the effective integration time $t_{\rm obs}\times N_{\rm eff}$, which controls the instrumental noise level through $\sigma_{\rm rms}\propto (t_{\rm obs}N_{\rm eff})^{-1/2}$. The horizontal dashed line indicates $\sqrt{\z}=5$, corresponding to a nominal $5\sigma$ detection of inconsistency with the null hypothesis $\Q_2=1$ or $P_{\rm line}^{\rm(B19)}=P_{\rm line}^{\rm true}$.

Several trends are apparent. First, at fixed statistic and redshift, the significance increases monotonically with $t_{\rm obs}N_{\rm eff}$. For the shortest effective exposures (dark curves), all three diagnostics remain below the $5\sigma$ threshold across the full redshift range, implying that neither $\Q_2$ nor the \B19\ power spectra are sensitive enough to detect the breakdown of the underlying assumptions. As the integration time is increased to $t_{\rm obs}N_{\rm eff}\sim10^{4}$ to $10^{5}$, both $\sqrt{\z_{\Q_2}}$ and $\sqrt{\z_{{[\textrm{C}\,\textsc{ii}]}}}$ cross the $5\sigma$ line at low redshift ($z\lesssim3$), and for the deepest configuration ($t_{\rm obs}N_{\rm eff}\sim10^{6}$) they reach very high significances, $\sqrt{\z}\gg10$, over a wide redshift range. This shows that, once instrumental noise is sufficiently suppressed, even modest fractional deviations from unity in $\Q_2$ or from the true spectrum in \B19\ become highly statistically significant.

Second, all three panels exhibit a strong redshift dependence: at fixed integration time, the significances decrease rapidly with increasing $z$. This reflects both the intrinsic dimming of the lines and the increasing dominance of noise at high redshift in our \texttt{LIMpy} modelling. For $\Q_2$, the deep survey curves fall from $\sqrt{\z}\gtrsim10^2$ at $z\sim1$ to $\mathcal{O}(1)$ by $z\sim5$–6, beyond which the estimator becomes effectively consistent with unity within the available sensitivity. The \B19\ significances for \CII\ and \NII\ show analogous behaviour, though with notable differences in detail: because \CII\ is brighter, $\sqrt{\z_{{[\textrm{C}\,\textsc{ii}]}}}$ remains above the $5\sigma$ threshold out to higher redshifts than $\sqrt{\z_{{[\textrm{N}\,\textsc{ii}]}}}$ for the same $t_{\rm obs}N_{\rm eff}$, while the fainter \NII\ line requires the longest integrations to reach comparable sensitivity.

Taken together, this figure illustrates how the $\Q_2$ statistic and the \B19\ power-spectrum estimates respond to improvements in survey depth across cosmic time. At low redshift and for sufficiently long observations, all three diagnostics provide very strong detections of the violation of the \B19\ consistency relations, with $\Q_2$ achieving sensitivity comparable to the direct power-spectrum tests. At higher redshifts, however, instrumental noise and line dimming reduce the available information, and the significances for all three statistics drop below the few-$\sigma$ level, delineating the redshift range over which multi-line LIM data can robustly diagnose the breakdown of the \B19\ assumptions for this particular line combination.

\subsubsection{With 21cm signal}
\label{sec:with_21cm}

Next, we assess the possibility of replacing one of the FIR lines with the 21-cm signal.
We simulate the 21-cm signal at \( z \leq 6 \) using a semi-numerical approach
\cite{Bagla:2009jy,Sarkar:2016lvb,Villaescusa-Navarro:2018vsg,Sarkar:2019ojl}. Halo catalogs from the \textsc{TNG300} simulation are used as the basis for assigning \HI\ content via the analytic prescription proposed in Ref.~\cite{Villaescusa-Navarro:2018vsg}. The neutral hydrogen mass associated with each halo of mass \( M_h \) is given by
\begin{equation}
M_{\rm HI}(M_h) = M_0 \left(\frac{M_h}{M_{\rm min,HI}}\right)^{\alpha} 
\exp\left[ -\left(\frac{M_{\rm min,HI}}{M_h}\right)^{0.35} \right],
\label{eq:HI_pop}
\end{equation}
where \( \alpha \), \( M_0 \), and \( M_{\rm min,HI} \) are redshift-dependent parameters calibrated from hydrodynamic simulations (see Table~I of Ref.~\cite{Villaescusa-Navarro:2018vsg}). The \HI\ overdensity field \( \delta_{\rm HI} \) is computed on a grid and converted into brightness temperature fluctuations using
\begin{equation}
\delta_{\rm 21-cm}(\mathbf{k},z) = T_b(z) \delta_{\rm HI}(\mathbf{k},z)\,{\rm mK},
\end{equation}
with the mean brightness temperature given by
\begin{equation}
T_b(z) = 189h \frac{H_0(1+z)^2}{H(z)}\Omega_{\rm HI}(z)\,{\rm mK},
\end{equation}
where \( \Omega_{\rm HI}(z) \) is the cosmological \HI\ density parameter, and \( H(z) \) and \( H_0 \) are the Hubble parameter at redshift \( z \) and at the present epoch, respectively.

\begin{figure*}
    \centering
    \includegraphics[width=1\linewidth]{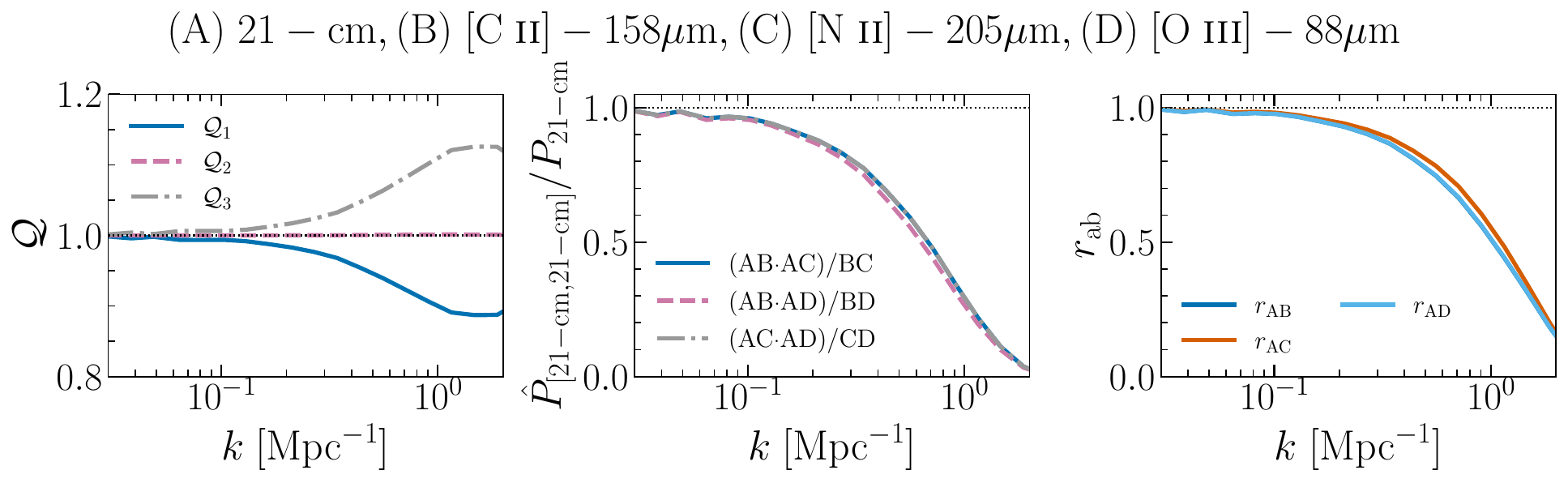}
    \caption{Noiseless performance of the $\Q$– and \B19–estimators when one of the four tracers is the post–reionization 21–cm field at $z=1$, combined with three FIR lines (\CII, \NII, and \OIII). 
The left panel shows $\Q_1$, $\Q_2$, and $\Q_3$ as functions of $k$. 
All three combinations are consistent with unity on large scales, indicating that 21–cm and the FIR lines trace the same underlying density field there, but $\Q_1$ and $\Q_3$ deviate by $\sim10\%$ at $k\sim1\,\mpci$, while $\Q_2$ remains nearly flat over the full range. 
The middle panel displays the ratio of the \B19–reconstructed 21–cm power spectrum to the true spectrum for different tri–line combinations; the estimator is unbiased on linear scales but increasingly underestimates the 21–cm power once $k\gtrsim0.1$–$0.2\,\mpci$. 
The right panel shows the cross–correlation coefficients between the 21–cm field and each FIR tracer, which are close to unity at small $k$ but decline rapidly towards zero on non–linear scales. 
The simultaneous loss of correlation and the breakdown of both $\Q$ and \B19\ at high $k$ highlight the increased vulnerability of the estimator when 21–cm is included as a tracer.
}
    \label{fig:Q_no_noise_21}
\end{figure*}

Figure~\ref{fig:Q_no_noise_21} shows the corresponding noiseless test 
with three star-formation–tracing FIR lines \CII\, \NII\, and \OIII\, and the 
21-cm field at $z=1$. The left panel displays the three $\Q_i$ combinations defined above. On large scales all three remain very close to unity, as in the purely FIR case, indicating that the 21-cm field traces the same large-scale structure as the FIR lines. On smaller, mildly non-linear scales the behaviour diverges: $\Q_1$ drifts to values $\sim 10\%$ below unity by $k\sim1,\mpci$, while $\Q_3$ rises to $\sim 10$ to 15\% above unity over the same range. Remarkably, $\Q_2$ remains almost perfectly flat, with deviations well below the percent level across the entire $k$-interval. This pattern shows that, once the 21-cm field is included, different $\Q_i$ combinations respond very differently to the mismatch in small-scale clustering between \HI\ and the FIR tracers, with $\Q_2$ effectively “protected’’ against these differences whereas $\Q_1$ and $\Q_3$ pick up systematic departures.
The middle panel examines how this impacts the \B19\ reconstruction of the 21-cm\ power spectrum. The recovered $P_{\rm 21-cm}$ agrees with the true spectrum only on large scales; the ratio $\hat{P}_{[\rm 21-cm,\rm 21-cm]}/P_{\rm 21-cm}$ begins to fall below unity already around $k\simeq0.1$–$0.2,\mpci$ and rapidly declines at higher $k$, dropping to $\mathcal{O}(0.1)$ by $k\sim1\,\mpci$ almost independently of the particular tri-line combination used. Compared to the four–FIR-line case, this demonstrates that using 21-cm as one of the inputs makes the \B19\ estimator substantially more vulnerable to small-scale effects like the non-linear bias, leading to a strong underestimation of the 21-cm\ power spectrum on quasi-linear and non-linear scales.
The right panel clarifies the origin of these trends by showing the cross-correlation coefficients between 21-cm and each FIR line. All three correlations are essentially unity at the largest scales, but they decline steadily beyond $k\sim0.1,\mpci$ and approach zero near $k\sim1$–$2,\mpci$. Thus, while \HI\ and the FIR tracers are nearly perfectly correlated on linear scales, they decorrelate rapidly in the non-linear regime, reflecting their different halo weighting and astrophysical dependence. The scale at which this decorrelation sets in coincides with the onset of the departures seen in $\Q_1$, $\Q_3$, and in the \B19\ reconstruction of $P_{\rm 21-cm}$.

\begin{figure*}
    \centering
    \includegraphics[width=0.77\linewidth]{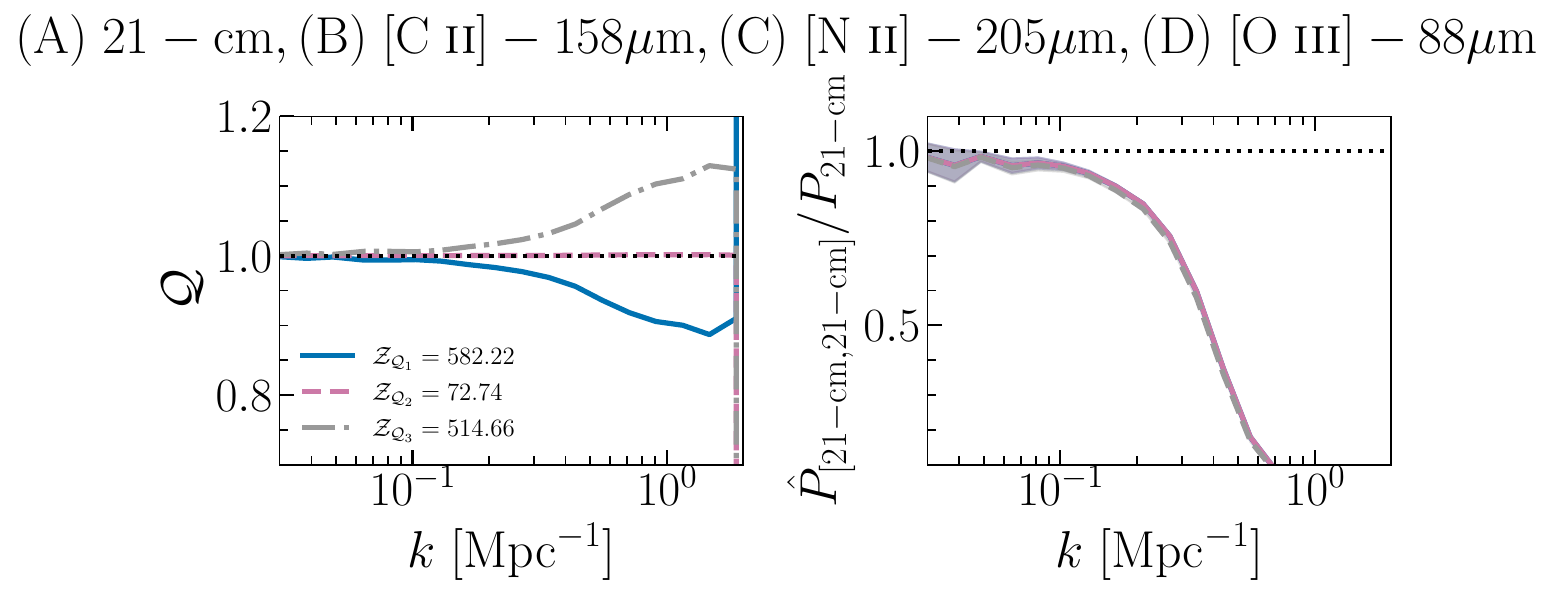}
    \caption{Same as Fig.~\ref{fig:Q_no_noise_21}, but including instrumental noise for both the LIM and 21–cm surveys. 
For the FIR tracers we assume the fiducial Super–LIM noise level with $N_{\rm det}t_{\rm obs}=2\times10^5\,{\rm hr}$, while for the 21–cm line we adopt a CHORD like survey with the observation time $t_{\rm obs,21-cm}=2000\,{\rm hr}$. 
The left panel shows the noisy measurements of $\Q_1$, $\Q_2$, and $\Q_3$ versus $k$, with shaded bands indicating the scatter across noise realizations; the legend reports the $\mathcal{Z}_{\Q_i}$ values quantifying the significance of the departure from unity. 
Even in the presence of noise, all three $\Q_i$ combinations remain consistent with $\Q=1$ on the largest scales but deviate with high significance on quasi–linear and non–linear scales, with $\Q_1$ and $\Q_3$ showing of order $10\%$ departures that are many sigma significant. 
The right panel shows the ratio of the \B19–reconstructed 21–cm power spectrum to the true spectrum: it is unbiased at $k\lesssim0.1\,\mpci$, but strongly suppressed at higher $k$, with the bias greatly exceeding the statistical uncertainties. 
The close correspondence between the scales where $\Q_i$ drifts from unity and where $\hat P_{\rm 21cm}^{\B19}$ becomes biased demonstrates that $\Q$ remains an effective null test even with realistic noise.
}
    \label{fig:Q_noise_21}
\end{figure*}

Figure~\ref{fig:Q_noise_21} shows the corresponding results when realistic instrumental noise is included for the mixed 21-cm--FIR configuration. For the LIM tracers we adopt the \textsc{Super-LIM} setup with an effective exposure of \(N_{\rm det} t_{\rm obs} = 2\times10^{5}\,{\rm hr}\).
For the 21-cm observations we adopt a CHORD-like rectangular array configuration and an integration time of ($t_{\rm obs} = 2000\,{\rm hr}$). The details of the noise-variance calculation are given in Section~\ref{sec:21cm_noise}. For each line we generate 100 independent noise realizations, and the means and variances shown in the figure are estimated from this ensemble.

The left panel displays the three \(\Q_i\) combinations, with shaded bands indicating the scatter across noise realizations and the legend quoting the corresponding  \(\mathcal{Z}_{\Q_i}\) values.
Despite the added noise, the large-scale behaviour of \(\Q_i\) closely resembles the noiseless case: all three combinations remain very near unity for \(k \lesssim 0.1\,\mpci\), confirming that the 21-cm field and the FIR lines still trace the same long-wavelength modes. On smaller scales the systematic trends seen previously are clearly preserved and become highly significant. Both \(\Q_1\) and \(\Q_3\) drift away from unity by \(\sim 10\%\) over \(k \sim 0.3\)--\(1\,\mpci\), while \(\Q_2\) remains much flatter, deviating only at the few-per-cent level. The quoted \(\mathcal{Z}_{\Q_i}\) values show that all three combinations detect the breakdown of the underlying assumptions with overwhelming significance: \(\Q_2\) already yields \(\mathcal{Z}_{\Q_2} \simeq 73\), and \(\Q_1\) and \(\Q_3\) are even more discrepant. The increase of the shaded bands at the largest \(k\) reflects the growing impact of thermal noise, but this additional variance is small compared to the systematic shift in the means, so the deviations remain very strongly detected.

The right panel demonstrates that the \B19\ reconstruction of the 21-cm auto-power spectrum is likewise reliable only on large scales. For \(k \lesssim 0.1\,\mpci\) the ratio \(\hat{P}_{[\mathrm{21-cm,\mathrm{21-cm}]}}/P_{\mathrm{21-cm}}\) is statistically consistent with unity, but it falls rapidly once quasi-linear scales are included, dropping below \(50\%\) by \(k \sim 0.3\,\mpci\) and approaching zero at \(k \gtrsim 1\,\mpci\). The shaded uncertainty bands again grow towards high \(k\), reflecting the combination of 21-cm thermal noise and the loss of correlation with the FIR tracers, yet the systematic suppression of power is much larger than the statistical errors. Comparing the two panels, we see that the scales where \(\hat{P}_{[\mathrm{21-cm},\mathrm{21-cm}]}\) becomes strongly biased coincide with those where the \(\Q_i\) combinations depart from unity, showing that even in the presence of realistic instrumental noise the \(\Q\)-estimator remains a powerful null test for diagnosing the failure of the \B19\ assumptions when a 21-cm tracer is included.

\begin{figure*}
    \centering
    \includegraphics[width=0.77\linewidth]{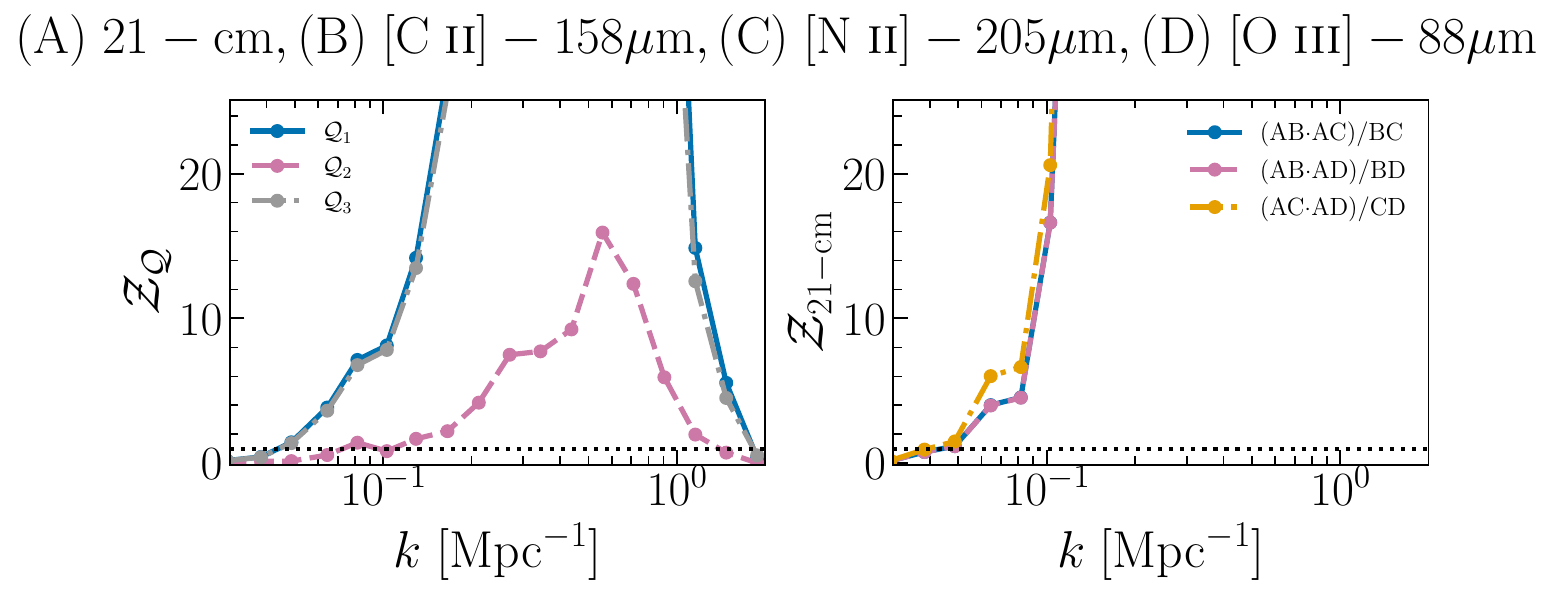}
    \caption{Per–bin significance of the departures from the \B19\ consistency relations for the mixed 21–cm–plus–FIR configuration with Super–LIM depth $N_{\rm det}t_{\rm obs}=2\times10^5\,{\rm hr}$ and $t_{\rm obs,21-cm}=2000\,{\rm hr}$. 
Left: $\mathcal{Z}_{\Q}(k)$ for the three combinations $\Q_1$, $\Q_2$, and $\Q_3$. 
All three are consistent with unity at very large scales, but $\Q_1$ and $\Q_3$ rapidly exceed the $1\sigma$ threshold for $k\gtrsim0.05\,\mpci$, reaching $\mathcal{Z}_\Q\gtrsim20$ at mildly non–linear scales, while $\Q_2$ remains more conservative and only attains a few–$\sigma$ significance at its peak. 
Right: $\mathcal{Z}(k)$ for the \B19\ reconstruction of the 21–cm power spectrum, computed for each tri–line combination entering the estimator. 
All combinations show a very rapid rise in significance once non–linear scales are included, surpassing $5\sigma$ already by $k\simeq0.08\,\mpci$ and reaching very large Z–values by $k\sim0.2\,\mpci$. 
The matching turnover scales in the two panels confirm that the $\Q$–estimator provides a sensitive, scale–resolved null test for the validity of the \B19\ relations in the presence of a 21–cm tracer.
}
    \label{fig:Z_k_21}
\end{figure*}

Figure~\ref{fig:Z_k_21} shows the scale-by-scale $\mathcal{Z}$-statistics for the mixed 21-cm–FIR configuration, using the quantity \(\mathcal{Z}_X(k)\) defined previously in Eq.~\eqref{eq:Zk}. The left panel displays \(\mathcal{Z}_{\Q}(k)\) for the three \(\Q_i\) combinations. On the largest scales (\(k \lesssim 0.05\,\mpci\)) all three curves lie close to the \(1\sigma\) threshold (horizontal dotted line), indicating that the \(\Q_i\) remain statistically consistent with unity once noise and sample variance are included. At slightly smaller scales, however, \(\Q_1\) and \(\Q_3\) begin to respond very strongly to the mismatch between the 21-cm and FIR fields: their $\mathcal{Z}$-values grow steeply with \(k\), reaching \(\mathcal{Z}_{\Q} \gtrsim 20\) over the range \(k \sim 0.1\)–\(0.2\,\mpci\). In contrast, \(\Q_2\) remains comparatively conservative, exceeding the \(1\sigma\) level only around \(k \sim 0.1\,\mpci\) and peaking at more modest values of a few–\(\sigma\) before declining again at the highest \(k\), where the errors become large.

Note that, the rise and subsequent fall of the \(\mathcal{Z}_{\Q}(k)\) curves at intermediate and high $k$ is primarily driven by the $k$-dependence of 
the 21\,cm instrumental noise. 
Unlike the star-formation lines, whose noise is approximately white over the 
scales of interest, the 21\,cm noise increases rapidly with $k$. 
As a result, cross-power spectra involving the 21\,cm field become 
progressively more noise-dominated at high $k$, 
producing the observed turnover.

The right panel shows the corresponding $\mathcal{Z}$-statistics for the \B19\ reconstruction of the 21-cm power spectrum, \(\mathcal{Z}_{\mathrm{21-cm}}(k)\), for the three tri-line combinations used in the estimator. All combinations behave almost identically: they are consistent with unity at the largest scales, but their $\mathcal{Z}$-values rise very rapidly once non-linear scales are included, surpassing the \(5\sigma\) level by \(k \sim 0.08\,\mpci\) and reaching \(\mathcal{Z}_{\mathrm{21-cm}} \gtrsim 20\) at slightly higher \(k\). The close correspondence between the scales at which \(\mathcal{Z}_{\Q}(k)\) and \(\mathcal{Z}_{\mathrm{21-cm}}(k)\) first exceed unity confirms that the \(\Q\)-estimator provides a sensitive, per-bin null test for diagnosing the breakdown of the \B19\ assumptions in the presence of a 21-cm tracer.

\begin{figure*}
    \centering
    \includegraphics[width=1\linewidth]{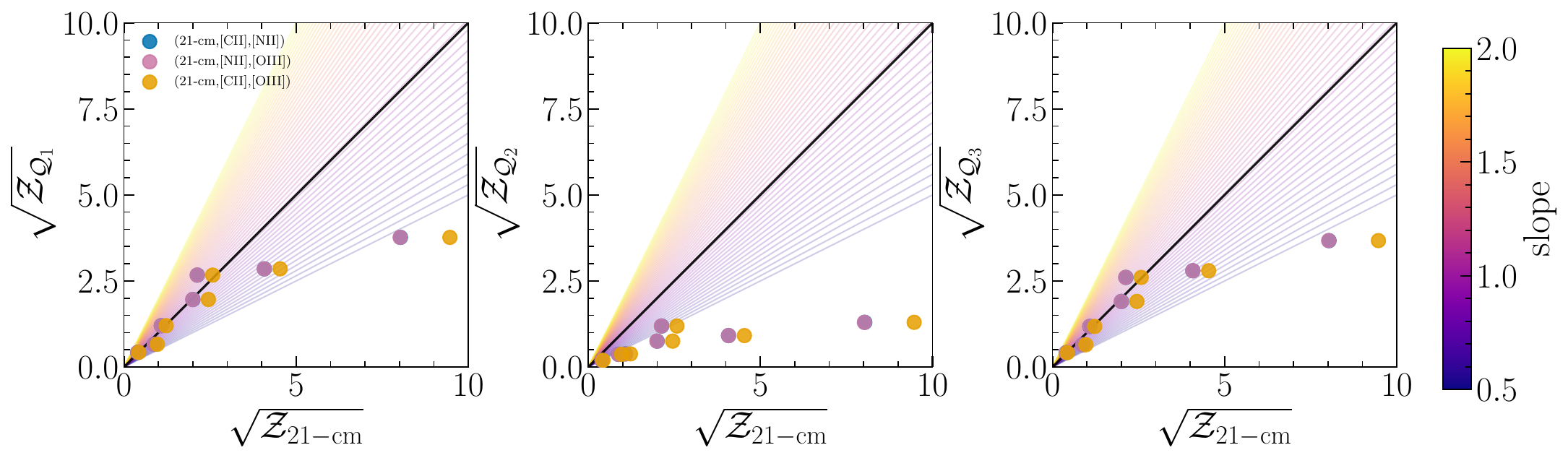}
    \caption{Comparison of the $\Q$–based and 21–cm \B19–based significances for the mixed 21–cm–plus–FIR configuration, in the same format as Fig.~\ref{fig:scatter_LIM_1000}. 
In each panel the horizontal axis shows $\sqrt{\z_{\rm 21-cm}}$, the global significance of the deviation between the \B19–reconstructed and true 21–cm power spectra, while the vertical axis shows $\sqrt{\z_{\Q_i}}$ for $i=1,2,3$ (left to right). 
Points correspond to different choices of the two FIR companions to the 21–cm field and are colour–coded by the tri–line combination. 
For $\Q_1$ and $\Q_3$ most points lie along slopes of order unity, indicating that these combinations are almost as sensitive as the 21–cm \B19\ estimator in diagnosing the breakdown of the underlying assumptions, with some configurations yielding comparable significances. 
By contrast, $\Q_2$ is visibly more conservative, with most points lying at slopes $m\lesssim0.5$, consistent with its reduced per–bin sensitivity in Fig.~\ref{fig:Z_k_21}. 
Taken together, the panels show that while the 21–cm \B19\ estimator typically provides the strongest constraint, the $\Q_i$ statistics remain highly correlated with it and offer complementary, noise–robust consistency checks.
}
    \label{fig:Z_21_scatter}
\end{figure*}

Figure~\ref{fig:Z_21_scatter} shows the detection significances in the same format as Figure~\ref{fig:Z_k_21}. For each tri-line configuration we plot \(\sqrt{ \mathcal{Z}_{\Q_i}}\) against \(\sqrt{ \mathcal{Z}_{\rm 21-cm}}\), where the latter quantifies the deviation of the \B19\ 21-cm power-spectrum estimate from the true spectrum. The three panels correspond to \(\Q_1\), \(\Q_2\), and \(\Q_3\) (left to right). Each point represents one choice of FIR companions to the 21-cm field, with colours indicating whether the triad is (21-cm, [C\,\textsc{ii}], [N\,\textsc{ii}]), (21-cm, [N\,\textsc{ii}], [O\,\textsc{iii}]), or (21-cm, [C\,\textsc{ii}], [O\,\textsc{iii}]). As before, the coloured fans of straight lines mark constant slopes \(m\), such that a point on a given line satisfies \(\sqrt{ \mathcal{Z}_{\Q_i}} = m \sqrt{ \mathcal{Z}_{\rm 21-cm}}\); the black diagonal corresponds to \(m=1\).

In the \(\Q_1\) panel, most points cluster around slopes of order unity, indicating that the global significance of the departures inferred from \(\Q_1\) roughly tracks that from the 21-cm \B19\ estimator. A subset of configurations, particularly those involving the brighter [C\,\textsc{ii}] and [O\,\textsc{iii}] lines, lie close to or slightly above the \(m=1\) line, showing that \(\Q_1\) can be as sensitive as the direct comparison of \(\hat{P}_{[\rm 21-cm,\rm 21-cm]}\) with the true power spectrum. Others fall below the diagonal, reflecting cases where the \B19\ reconstruction of the 21-cm spectrum is more strongly inconsistent than indicated by the corresponding \(\Q_1\) combination.

The middle panel demonstrates that \(\Q_2\) is noticeably more conservative. Almost all points lie beneath the \(m=1\) line and in the region \(m \lesssim 0.5\), implying that, when integrated over all \(k\)-modes, \(\Q_2\) generally yields a lower  significance than \(\mathcal{Z}_{\rm 21-cm}\). This behaviour is consistent with the scale-by-scale analysis, where \(\Q_2\) remained closer to unity than \(\Q_1\) and \(\Q_3\) even on scales where the 21-cm \B19\ estimator was strongly biased.

In contrast, the \(\Q_3\) panel (right) resembles \(\Q_1\) more closely: several points again lie near slopes \(m \sim 1\), particularly for configurations that include [C\,\textsc{ii}] and [O\,\textsc{iii}], while others sit at lower slopes where the \B19\ departures are more significant.

\begin{figure*}
    \centering
    \includegraphics[width=0.77\linewidth]{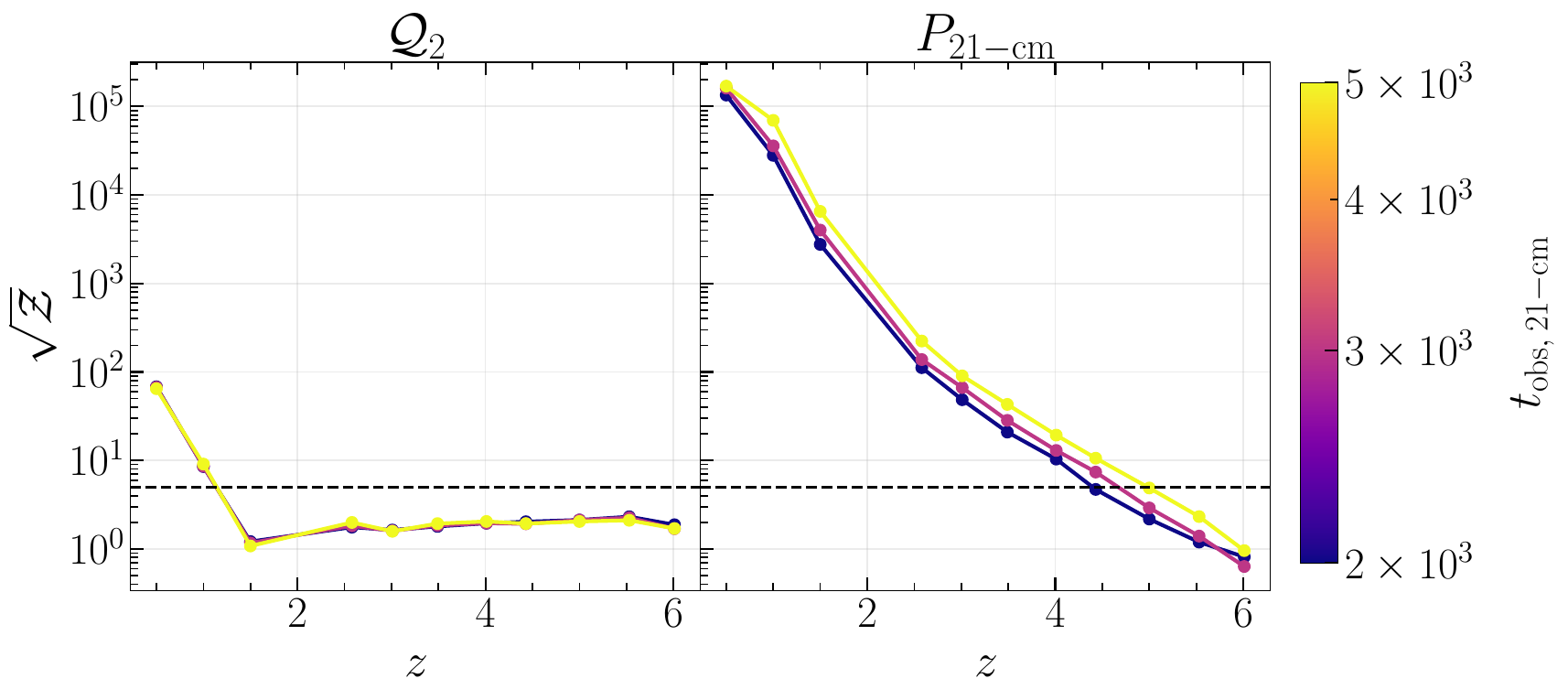}
    \caption{Redshift dependence of the detection significance when varying only the 21–cm observing time while keeping the LIM depth fixed at $N_{\rm eff}t_{\rm obs}=2\times10^5\,{\rm hr}$. 
The left panel shows $\sqrt{\z_{\Q_2}}$ and the right panel $\sqrt{\z_{P_{\rm 21cm}}}$, where $\z_{\Q_2}$ quantifies the departure of $\Q_2$ from unity and $\z_{{\rm 21-cm}}$ that of the \B19–reconstructed 21–cm power spectrum from the true spectrum. 
Coloured curves correspond to different 21–cm integrations $t_{\rm obs,21-cm}$ (colour scale), and the horizontal dashed line indicates a nominal $5\sigma$ threshold. 
At low redshift, both diagnostics are extremely significant, with $\sqrt{\z}$ values far above $5$, but their behaviour diverges at higher $z$: $\sqrt{\z_{\Q_2}}$ drops rapidly and saturates at ${\cal O}(1)$ for $z\gtrsim1$ almost independently of $t_{\rm obs,21-cm}$, whereas $\sqrt{\z_{{\rm 21cm}}}$ remains above $5\sigma$ out to $z\sim4$–5 and continues to increase with longer integrations. 
This illustrates that, once the LIM depth is fixed, increasing 21–cm observing time primarily enhances the constraining power of the direct \B19\ 21–cm power–spectrum test, while the $\Q_2$ statistic at $z\gtrsim1$ is limited by LIM noise rather than by 21–cm thermal noise.
}
    \label{fig:Z_tobs_21cm}
\end{figure*}

Figure~\ref{fig:Z_tobs_21cm} summarizes how the detection significance changes with redshift when we vary only the 21-cm observation time, keeping the Super-LIM depth fixed at \(N_{\rm eff} t_{\rm obs} = 2\times10^{5}\,{\rm hr}\). The colour bar encodes 
the 21-cm observation time \(t_{\rm obs,21cm}\), with darker blue colours corresponding to shorter integrations and lighter yellow colours to longer ones. As before, we plot \(\sqrt{\mathcal{Z}_X}\) which is defined in Eq.~\eqref{eq:chi_def}, and the horizontal dashed line marks \(\sqrt{\mathcal{Z}}=5\), our nominal \(5\sigma\) threshold.

At the lowest redshift bin the deviations from the null hypothesis are extremely significant, with \(\sqrt{\mathcal{Z}_{\Q_2}}\sim{\cal O}(10^2)\), but the significance drops rapidly with increasing \(z\). By \(z\simeq1.5\) the curves for all three 21-cm observing times have fallen to \(\sqrt{\mathcal{Z}_{\Q_2}}\sim1\), and remain at the \({\cal O}(1)\) level up to \(z\simeq6\). The weak dependence on \(t_{\rm obs,21cm}\) at \(z\gtrsim1\) indicates that, once the 21-cm experiment reaches a few \(10^{3}\)~hr of integration, the variance of \(\Q_2\) is dominated by LIM noise and by the intrinsic decorrelation between 21-cm and FIR tracers rather than by 21-cm thermal noise.

The right panel displays the corresponding results for the \B19\ reconstruction of the 21-cm auto-power spectrum, \(\sqrt{\mathcal{Z}_{{\mathrm{21-cm}}}}\). Here the behaviour is markedly different: for all three observing times the significance is enormous at low redshift, \(\sqrt{\mathcal{Z}_{{\mathrm{21-cm}}}}\gtrsim10^{4}\) at \(z\lesssim1\), and then declines roughly exponentially with \(z\). Even so, the \B19\ estimator remains well above the \(5\sigma\) threshold out to \(z\simeq4\)–5, with longer 21-cm integrations systematically yielding higher significances at fixed redshift. Only at the highest redshifts does \(\sqrt{\mathcal{Z}_{{\mathrm{21-cm}}}}\) approach the few–\(\sigma\) regime. 

Taken together, these trends show that, with the LIM depth held fixed, increasing \(t_{\rm obs,21cm}\) primarily boosts the constraining power of the direct \B19\ 21-cm power-spectrum test over a wide redshift range, while the \(\Q_2\) statistic is extremely sensitive at low redshift but quickly saturates to \({\cal O}(1)\) significance at \(z \gtrsim 1\).

\begin{figure*}
    \centering
    \includegraphics[width=0.77\linewidth]{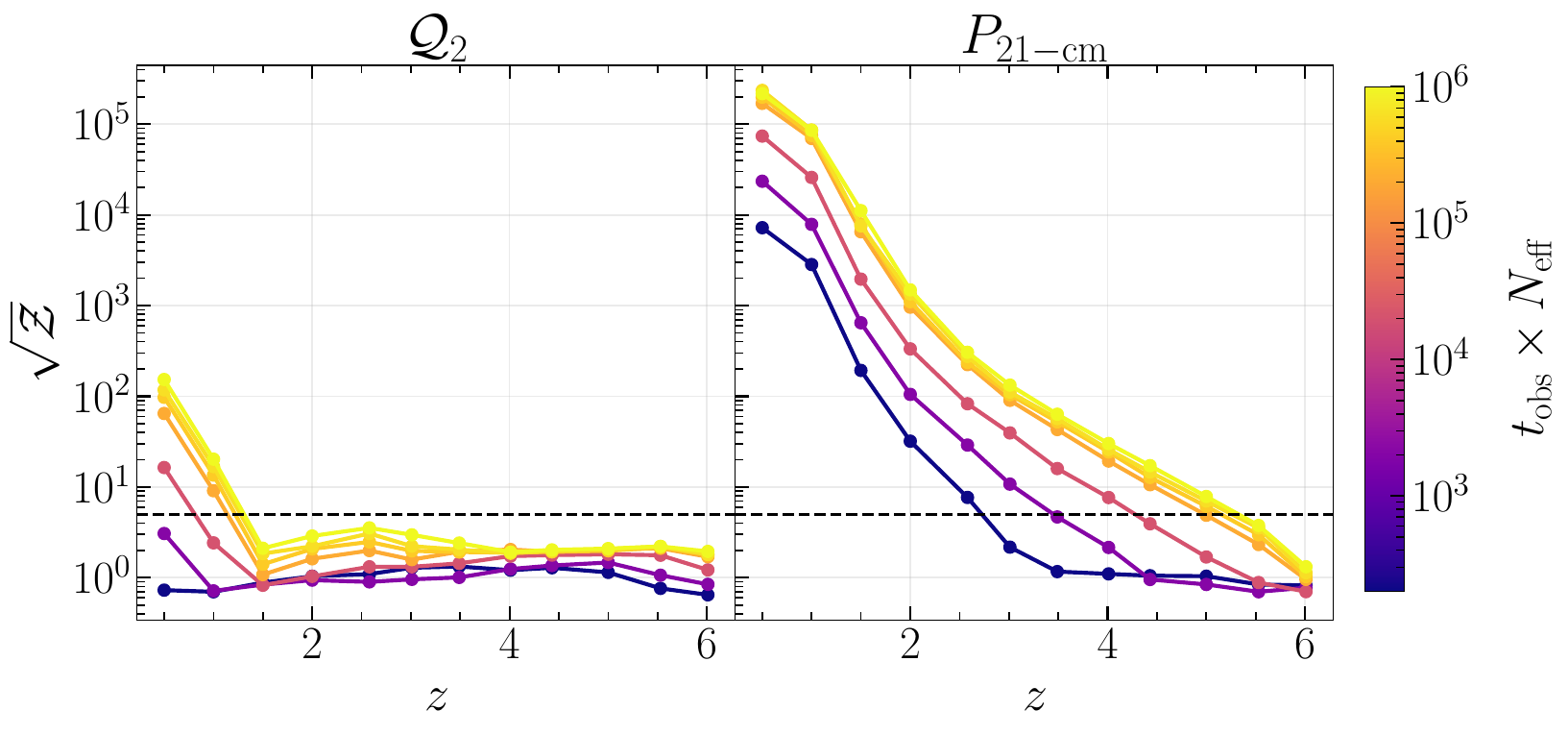}
    \caption{Complementary to Fig.~\ref{fig:Z_tobs_21cm}: here we fix the 21–cm observing time to $t_{\rm obs,21-cm}=5000\,{\rm hr}$ and vary only the LIM depth, parameterized by $t_{\rm obs}N_{\rm eff}$ (colour scale). 
The left panel shows $\sqrt{\z_{\Q_2}}$ as a function of redshift, while the right panel shows $\sqrt{\z_{{\rm 21cm}}}$ for the \B19–reconstructed 21–cm power spectrum. 
At low redshift, increasing the LIM depth boosts the significance of both diagnostics, with $\sqrt{\z_{\Q_2}}$ and $\sqrt{\z_{{\rm 21cm}}}$ rapidly exceeding the $5\sigma$ threshold (horizontal dashed line). 
At $z\gtrsim1.5$, however, $\sqrt{\z_{\Q_2}}$ converges to values of order unity with only a weak dependence on $t_{\rm obs}N_{\rm eff}$, indicating that the variance of $\Q_2$ is then dominated by 21–cm noise rather than by LIM sensitivity. 
By contrast, $\sqrt{\z_{{\rm 21cm}}}$ continues to grow appreciably with LIM depth at all redshifts, as deeper LIM observations tighten the cross–power measurements entering the \B19\ estimator. 
Together with Fig.~\ref{fig:Z_tobs_21cm}, this figure highlights the complementary roles of the LIM and 21–cm experiments in setting the overall sensitivity of $\Q_2$ and of the 21–cm \B19\ power–spectrum reconstruction.
}
    \label{fig:Z_tobs_LIM}
\end{figure*}

Figure~\ref{fig:Z_tobs_LIM} presents the complementary case to
Figure~\ref{fig:Z_tobs_21cm}. Here we fix the 21-cm observing time to
\(t_{\rm obs,21cm}=5000\,{\rm hr}\) and vary only the depth of the
Super-LIM experiment, parameterized by \(t_{\rm obs} \times N_{\rm eff}\)
(colour-coded on the right-hand side). 

The left panel displays the \(\Q_2\) statistic. At the lowest redshift the significance is
large, \(\sqrt{\mathcal{Z}_{\Q_2}}\sim{\cal O}(10)\), for the modest value 
\(t_{\rm obs} \times N_{\rm eff}=10^4\,{\rm hrs}\) and
increases further with LIM depth, reflecting the fact that both the 21-cm
and FIR tracers are measured at very high signal-to-noise. Between
\(z\simeq1.5\) and \(z\simeq6\), however, all curves converge to
\(\sqrt{\mathcal{Z}_{\Q_2}}\sim1\)–3 with only a weak dependence on
\(t_{\rm obs} N_{\rm eff}\). In this regime the variance of \(\Q_2\) is
possibly dominated by the 21-cm noise, so further improving the LIM sensitivity
has only a modest impact on the overall significance.

The right panel shows the corresponding behaviour for the \B19\
reconstruction of the 21-cm auto-power spectrum. We see that, varying \(t_{\rm obs} N_{\rm eff}\) produces a still monotonic, increase in \(\sqrt{\mathcal{Z}_{{\rm 21-cm}}}\) at all
redshifts: deeper LIM data tighten the constraints on the cross-spectra
entering the estimator and enhance the significance with which the
systematic departure of \(\hat{P}_{[\rm 21-cm,21-cm]}\) from truth is detected. Even for
the shallowest LIM configuration the 21-cm power-spectrum deviations are
highly significant at low redshift, while for the deepest LIM depths the
estimator remains well above the \(5\sigma\) level out to \(z\sim5\).

Comparing this figure to Figure~\ref{fig:Z_tobs_21cm} highlights the
complementary roles of the two experiments. Increasing the 21-cm
observing time primarily boosts the sensitivity of the direct
\(\hat{P}_{[\rm 21-cm,21-cm]}\) test, with relatively little effect on
\(\Q_2\) at \(z\gtrsim1\), whereas increasing the LIM depth strengthens
both diagnostics but with a more pronounced impact on the 21-cm
power spectrum significance than on the already noise-saturated \(\Q_2\)
statistic at high redshift.

\section{Summary and Discussion}
\label{sec:summary}

In this work we have introduced and tested a new statistic, $\Q$, designed as a
data–driven null test for multi–line intensity mapping analyses that use the
\B19\ cross–spectrum estimator to reconstruct auto–power spectra. By forming
ratios of cross–power spectra between four tracers, $\Q$ cancels the explicit
dependence on the matter power spectrum and on their linear bias amplitudes.
When all tracers are well described as linearly biased, highly correlated
probes of the same underlying field, $\Q$ is expected to be unity with very
small variance; significant departures from unity signal a breakdown of these
assumptions.

We have shown, using a combination of idealized Gaussian fields and more realistic LIM/
21-cm maps built from halo catalogs, that this intuition carries over to
practical applications. In regimes where all lines are strongly cross–correlated and dominated by large, linear
scales, the \B19\ reconstructions are unbiased and the different $\Q_i$
combinations cluster tightly around $\Q=1$. As soon as tracer decorrelation
sets in---because of changes in halo weighting, non–linear bias, or differing
astrophysics---the \B19\ estimates become biased and $\Q$ exhibits clear,
scale–dependent deviations from unity. This behaviour persists in the presence
of realistic instrumental noise: on large scales $\Q$ remains consistent with
unity, while on smaller scales its significance tracks the scales where \B19\
reconstructions fail.

The main practical significance of these results is that $\Q$ provides an
internal, survey-level diagnostic for the reliability of \B19–based
auto–spectrum reconstructions, without requiring an external, foreground–free
measurement of the auto–power spectrum. In particular:
\begin{itemize}
  \item $\Q$ and \B19\ are tightly correlated diagnostics: the $k$- and
  redshift-ranges where $\Q$ departs from unity at high significance coincide
  with the regimes where \B19\ becomes biased. $\Q$ therefore offers a
  convenient way to define the scale cuts and redshift ranges over which
  \B19-based reconstructions can be trusted.

  \item Different $\Q$ combinations ($\Q_1,\Q_2,\Q_3$) and different choices of
  line triads in \B19\ have complementary sensitivity. No single combination is
  universally optimal; robust analyses should consider all available
  configurations jointly and look for consistent trends. In our tests, some
  $\Q_i$ are more conservative but low-variance probes of departures, while
  others are more sensitive to small non-linear effects.

  \item The diagnostic power of $\Q$ depends on survey depth, tracer choice, and
  redshift. For bright far–infrared lines at low redshift, $\Q$ reaches very
  high significance and can sharply delineate the reliable \B19\ regime. At
  higher redshift or for fainter tracers, $\Q$ becomes noise-limited and
  mainly constrains the largest scales. When 21\,cm is included as one of the
  tracers, large-scale sensitivity improves but small-scale decorrelation
  quickly limits the usefulness of \B19\ and is clearly flagged by $\Q$.
\end{itemize}

Taken together, these results suggest several directions for future work. A
natural next step is to include foregrounds and interloper contamination, and
to test how robust $\Q$ remains after realistic cleaning and masking. Recent
forecast studies of 21\,cm$\times$[C\,\textsc{ii}] cross–correlations have
shown explicitly how line interlopers and residual foregrounds can degrade or
bias LIM cross–power measurements, even when the underlying cosmological
fields remain in the linear regime (e.g.\ \cite{Fronenberg:2024olu}). In such
cases, emission from lines at different redshifts effectively adds extra,
partially correlated components to some of the observed maps, altering both
the relative amplitudes and the scale–dependence of the various cross–spectra.

Because $\Q$ is built from ratios of cross–spectra involving four tracers, it
is naturally sensitive to this kind of contamination. An interloper that
contributes strongly to only a subset of the lines will generically drive $\Q$
away from unity on scales where the true tracers alone would still follow the
linear–bias expectation. Conversely, if interlopers act predominantly as
uncorrelated noise for all of the relevant maps, their impact on $\Q$ should
largely average out. This suggests a concrete use of $\Q$ as a data–driven
interloper diagnostic: by comparing the behaviour of $\Q$ across different line
combinations, survey depths, and masking strategies, one can empirically
identify configurations in which residual interlopers are still present (for
example, when $\Q$ remains offset from unity even on the largest accessible
scales) and adjust the analysis accordingly. Configurations where $\Q$ is
consistent with unity across the linear regime provide direct support for the
common assumption that interloper contamination in cross–correlations can be
treated as effectively uncorrelated noise.

More sophisticated models for line emission and the 21\,cm signal (including
redshift–space distortions \cite{Sarkar:2018gcb,Sarkar:2019nak,Osinga:2025gpt} 
and feedback \cite{Chisari:2019tus,Medlock:2024wxu,Maraio:2024xjz,Miller:2025gwh}) will be required to translate
$\Q$–based diagnostics into quantitative priors on astrophysical parameters.
Incorporating explicit models for likely interloper populations within these
frameworks would allow deviations of $\Q$ from unity to be interpreted in terms
of the level, clustering, and redshift distribution of contaminant lines, thus
linking the null test more directly to survey and analysis choices. In
parallel, Fisher forecast–based survey design studies that include realistic
interlopers and foregrounds can be combined with $\Q$–based null tests: the
former quantify how assumed contamination levels propagate into parameter
constraints, while the latter provides an on–sky, survey–specific check of
whether those assumptions hold for a given dataset.

Finally, the formalism can be extended to additional tracers and to angular
power spectra, enabling $\Q$ to be applied directly to forthcoming multi–line
LIM and LIM–21\,cm cross-correlation data sets. In that role, $\Q$ can act as a
simple pre-analysis tool to identify the regions of $(k,z)$-space where the
data themselves support the assumptions underlying multi–tracer estimators such
as \B19, and where more flexible modeling will be required. With tools such as \B19, $\mathcal{Q}$, and new estimator proposals in the future, one opens the door to a truly multi-wavelength view of our intensity mapped Universe.

\begin{acknowledgments}
The authors acknowledge support from the Natural
Sciences and Engineering Research Council of Canada
through their Discovery Grants Program and their Alliance International Program, as well as
the William Dawson Scholar program at McGill University.
This research was enabled in part by support provided by Calcul Quebec\footnote{\href{https://www.calculquebec.ca/}{https://www.calculquebec.ca/}} and the Digital Research Alliance of Canada\footnote{\href{https://www.alliancecan.ca/}{https://www.alliancecan.ca/}}.
DS acknowledges the support of the Canada $150$ Chairs program, the Fonds de recherche du Qu\'{e}bec Nature et Technologies (FRQNT) and Natural Sciences and Engineering Research Council of Canada (NSERC) joint NOVA grant, and the Trottier Space Institute Postdoctoral Fellowship program.
The authors are grateful to the members of the `McGill Cosmic Dawn Group'---Hannah Fronenberg, Robert Pascua, Mike Wilensky, Matt\'eo Blamart, Kai-Feng Chen, Audrey Bernier, Franco Del Balso, Mariah Zeroug, Rebecca Ceppas de Castro, William Paty, Sophia Rubens, Kim Morel, Laurie Amen, Josh Goodeva, Andrei Li, Marek Deti\`ere-Venkatesh, and Arnab Chakraborty---for their helpful comments and suggestions at various stages of this work.
DS acknowledges helpful comments from Abigail Crites and Guochao Sun during discussions held while they were visiting McGill University.
DS acknowledges the organizers of the `Cosmology in Multicolour via Line Intensity Mapping Surveys' meeting held at the Institute for Fundamental Physics of the Universe (IFPU), Trieste, where a preliminary version of this work was presented, as well as useful comments from Andrei Mesinger, Anjan Ananda Sen, Anirban Roy, Azadeh Moradinezhad, Caroline Heneka, Mario Santos, Marta Spinelli, Raghunath Ghara, Suman Majumdar, and Tirthankar Roy Choudhury.

\end{acknowledgments}

\appendix

\section{Variance of $\mathcal{Q}(k)$}
\label{sec:variance}
In this Appendix, we derive theoretical expressions for the variance of $\mathcal{Q} \equiv [P_{ab}(k)\,P_{cd}(k)] /[{P_{ac}(k)\,P_{bd}(k)}] $. We stress, however, that these expressions are used only to guide our intuition, and that the error bars placed on $\mathcal{Q}$ in Section~\ref{sec:sf_lines} do not assume Gaussian errors because they are based on Monte Carlo simulations of noise realizations.

We consider four (complex) Fourier fields \(a,b,c,d\) measured in the same
\(k\)-bin with \(N_m(k)\) independent modes. Throughout, angle brackets denote
an ensemble average over realizations. We write
\begin{equation}
P_{xy}(k)\equiv \frac{1}{V} \langle x(\mathbf k)\,y^*(\mathbf k)\rangle
\end{equation}
and
\begin{equation}
N_{xy}(k)\equiv \frac{1}{V} \langle n_x(\mathbf k)\,n_y^*(\mathbf k)\rangle ,
\end{equation}
where \(N_{xy}\) is the instrumental noise
cross-power (usually \(N_{xy}=0\) for \(x\neq y\), and \(N_{xx}\equiv N_x\)).
The standard Gaussian covariance of two cross power spectra in the same bin is
\begin{align}
\mathrm{Cov}\!\big[\widehat P_{xy},\widehat P_{uv}\big]
 =&\frac{1}{N_m}
\Big[
\big(P_{xu}+N_{xu}\big)\big(P_{yv}+N_{yv}\big) \nonumber \\
&+\big(P_{xv}+N_{xv}\big)\big(P_{yu} +N_{yu}\big)
\Big].
\label{eq:GaussCov}
\end{align}
The corresponding variances follow by setting \((u,v)=(x,y)\).

Taking the natural logarithm of $\mathcal{Q}$ gives us
\begin{equation}
   \ln \mathcal{Q} \equiv\ln P_{ab}+\ln P_{cd}-\ln P_{ac}-\ln P_{bd},
\end{equation}
and to first order we have
\begin{equation}
\mathrm{Var}(\ln Q)=\sum_{X,Y}s_X s_Y\,\mathrm{Cov}(P_X,P_Y)/(P_X P_Y)\,,
\end{equation}
with \(X,Y\in\{ab,cd,ac,bd\}\) and signs
\(s_{ab}=s_{cd}=+1,\ s_{ac}=s_{bd}=-1\).
Since \(\mathrm{Var}(Q)/Q^2\simeq \mathrm{Var}(\ln Q)\) at this order, we obtain
\begin{align}
\frac{\mathrm{Var}\,Q}{Q^2}
&=\sum_{X}\frac{\mathrm{Var}(P_X)}{P_X^2}
+2\!\!\!\sum_{X<Y}\!\! s_X s_Y\,
\frac{\mathrm{Cov}(P_X,P_Y)}{P_X P_Y}.
\label{eq:master}
\end{align}
We now substitute the Gaussian covariances \eqref{eq:GaussCov} term by term.

\subsection*{Diagonal pieces}
\begin{align}
\frac{\mathrm{Var}(P_{ab})}{P_{ab}^2}
&=\frac{1}{N_m}\,
\frac{P_{ab}^2+(P_{aa}+N_a)(P_{bb}+N_b)}{P_{ab}^2},
\\
\frac{\mathrm{Var}(P_{cd})}{P_{cd}^2}
&=\frac{1}{N_m}\,
\frac{P_{cd}^2+(P_{cc}+N_c)(P_{dd}+N_d)}{P_{cd}^2},
\\
\frac{\mathrm{Var}(P_{ac})}{P_{ac}^2}
&=\frac{1}{N_m}\,
\frac{P_{ac}^2+(P_{aa}+N_a)(P_{cc}+N_c)}{P_{ac}^2},
\\
\frac{\mathrm{Var}(P_{bd})}{P_{bd}^2}
&=\frac{1}{N_m}\,
\frac{P_{bd}^2+(P_{bb}+N_b)(P_{dd}+N_d)}{P_{bd}^2}.
\end{align}

\subsection*{Off–diagonal pieces}
For compactness we list each covariance and its contribution to
\eqref{eq:master}.

\paragraph{(i) \( (ab,cd) \) with sign \(+\):}
\begin{align}
\frac{\mathrm{Cov}(P_{ab},P_{cd})}{P_{ab}P_{cd}}
&=\frac{1}{N_m}\,\frac{(P_{ac}P_{bd}+P_{ad}P_{bc})}
{P_{ab}P_{cd}}.
\end{align}

\paragraph{(ii) \( (ab,ac) \) with sign \( - \):}
\begin{align}
\frac{\mathrm{Cov}(P_{ab},P_{ac})}{P_{ab}P_{ac}}
&=\frac{1}{N_m}\,\frac{(P_{aa}+N_{aa})P_{bc}+(P_{ac}P_{ab})}
{P_{ab}P_{ac}}.
\end{align}

\paragraph{(iii) \( (ab,bd) \) with sign \( - \):}
\begin{align}
\frac{\mathrm{Cov}(P_{ab},P_{bd})}{P_{ab}P_{bd}}
&=\frac{1}{N_m}\,\frac{(P_{ab}P_{bd})+(P_{ad})(P_{bb}+N_b)}
{P_{ab}P_{bd}}.
\end{align}

\paragraph{(iv) \( (cd,ac) \) with sign \( - \):}
\begin{align}
\frac{\mathrm{Cov}(P_{cd},P_{ac})}{P_{cd}P_{ac}}
&=\frac{1}{N_m}\,\frac{(P_{ac}P_{cd})+(P_{cc}+N_c)P_{ad}}
{P_{cd}P_{ac}}.
\end{align}

\paragraph{(v) \( (cd,bd) \) with sign \( - \):}
\begin{align}
\frac{\mathrm{Cov}(P_{cd},P_{bd})}{P_{cd}P_{bd}}
&=\frac{1}{N_m}\,\frac{P_{bc}(P_{dd}+N_d)+(P_{cd}P_{bd})}
{P_{cd}P_{bd}}.
\end{align}

\paragraph{(vi) \( (ac,bd) \) with sign \( + \):}
\begin{align}
\frac{\mathrm{Cov}(P_{ac},P_{bd})}{P_{ac}P_{bd}}
&=\frac{1}{N_m}\,\frac{P_{ab}P_{cd}+P_{ad}P_{bc}}
{P_{ac}P_{bd}}.
\end{align}

\begin{widetext}
\subsection*{Collecting everything}
Inserting  the four diagonal and the six off–diagonal pieces into
\eqref{eq:master} and factoring out the common \(1/N_m\) yields

\begin{align}
\frac{\mathrm{Var}\,Q}{Q^2}
&=\frac{1}{N_m}\Bigg\{
\frac{P_{ab}^2+(P_{aa}+N_a)(P_{bb}+N_b)}{P_{ab}^2}
+\frac{P_{cd}^2+(P_{cc}+N_c)(P_{dd}+N_d)}{P_{cd}^2}
+\frac{P_{ac}^2+(P_{aa}+N_a)(P_{cc}+N_c)}{P_{ac}^2}
\nonumber\\
&\qquad\qquad\quad
+\frac{P_{bd}^2+(P_{bb}+N_b)(P_{dd}+N_d)}{P_{bd}^2}
+\,2\,\frac{(P_{ac}P_{bd}+P_{ad}P_{bc})}{P_{ab}P_{cd}}
-\,2\,\frac{(P_{aa}+N_{aa})(P_{bc})+(P_{ac}P_{ab})}{P_{ab}P_{ac}}
\nonumber\\
&\qquad\qquad\quad
-\,2\,\frac{(P_{ab}P_{bd})+(P_{ad})(P_{bb}+N_b)}{P_{ab}P_{bd}}
-\,2\,\frac{(P_{ac}P_{cd})+(P_{cc}+N_c)(P_{ad})}{P_{cd}P_{ac}}
-\,2\,\frac{(P_{bc})(P_{dd}+N_d)+(P_{cd}P_{bd})}{P_{cd}P_{bd}}
\nonumber\\
&\qquad\qquad\quad
+\,2\,\frac{(P_{ab}P_{cd}+P_{ad}P_{bc})}{P_{ac}P_{bd}}
\Bigg\}.
\label{eq:VarQ_full}
\end{align}
\end{widetext}
This expression guides our intuition in various limits:
\begin{itemize}
\item {Signal-only, independent noise (\(N_{xy}=0\) for \(x\neq y\)).}
Setting all auto-noise to \(N_x=0\) simplifies
\eqref{eq:VarQ_full} to the cosmic-variance piece. In the special case where
all four tracers are perfectly coherent with the same underlying modes
(\(P_{xy}^2=P_{xx}P_{yy}\) for all pairs), the bracket in
\eqref{eq:VarQ_full} vanishes and \(\mathrm{Var}(Q)\to 0\) at Gaussian order
(\emph{cosmic-variance cancellation}).

\item {Including auto-noise only.}
Replace \(P_{xx}\to P_{xx}+N_x\) wherever it appears in \eqref{eq:VarQ_full}.
This lowers the field-to-field coherences and sets the floor once CV cancels.

\item {Mode counting.}
All terms are proportional to \(1/N_m(k)\). For a 3D shell of width
\(\Delta k\), \(N_m(k)\propto V\,k^2\Delta k\); hence the absolute variance
falls roughly as \(k^{-2}\) (while the fractional error scales
\(\propto N_m^{-1/2}\)) until non-Gaussian or noise terms dominate.
\end{itemize}

\section{Instrument Noise for LIM experiments}
\label{sec: inst noise z6}

In Figure~\ref{fig:expts}, we show the frequency coverage of representative LIM surveys—including FYST, CONCERTO, EXCLAIM, TIME, COPSS, and COMAP—alongside the redshifted frequencies of key emission lines. Together, these surveys span a broad redshift range and enable cross-correlation studies across multiple tracers. In this Appendix, we describe how we model the instrumental noise for these observations, distinguishing between single-dish LIM experiments (targeting FIR lines) and interferometric 21-cm surveys. We also describe how this noise is incorporated into our simulated intensity maps.

\subsection{Far-Infrared Line Observations (Single-Dish)}
\label{sec:fir_noise}

\begin{table}[t]
\centering
\begin{tabular}{ll}
\hline\hline
\textbf{Parameter} & \textbf{Super-LIM (single dish)} \\
\hline
$\sigma_{\rm pix}$ (MJy/sr s$^{1/2}$)  & 0.21 \\
$N_{\rm det}$                         & 200 \\
$D_{\rm dish}$ (m)         & 12  \\
$\delta\nu$ (GHz)                   & 0.4 \\
$\Omega_{\rm survey}$ (deg$^2$)     & 100  \\
\hline\hline
\end{tabular}
\caption{Instrument parameters assumed for the Super-LIM (single-dish) experiment.}
\label{tab:superlim_params}
\end{table}

We model the instrumental noise for far-infrared (FIR) single-dish experiments targeting the \CII, \NII, \CI\ and \OIII\ emission lines. To generalize across multiple existing surveys, we define a notional instrument, \textbf{Super-LIM}, with specifications designed to reflect the combined capabilities of FYST, CONCERTO, EXCLAIM, and TIME. This hypothetical instrument is assumed to operate across a broad frequency range of $10$ to $1000$\,GHz. The assumed parameters for Super-LIM are summarized in Table~\ref{tab:superlim_params}.

Our noise modeling approach follows the framework developed in Ref.~\cite{Fronenberg:2024olu}, treating instrumental noise as a Gaussian random field with zero mean and root-mean-square (RMS) amplitude $\sigma_{\rm rms}$. This RMS noise level is related to the instrument configuration via the expression:
\begin{equation}
\sigma_{\rm rms} = \sigma_{\rm pix} \sqrt{\frac{\Omega_{\rm survey}}{N_{\rm det} \, t_{\rm survey} \, \Omega_{\rm pix}}},
\label{eq:rms-lim}
\end{equation}
where $\sigma_{\rm pix}$ is the instantaneous detector sensitivity per pixel, $N_{\rm det}$ is the number of spectrometers, $t_{\rm survey}$ is the total observing time, and $\Omega_{\rm survey}$ is the total surveyed area~\cite{Dumitru:2018tgh,Fronenberg:2023qtw}. The solid angle of a single beam, which defines the instrumental pixel size, is given by $\Omega_{\rm pix} = 2\pi \sigma_{\rm beam}^2$, where the beam width is
\begin{equation}
\sigma_{\rm beam} = \left(\frac{1.22 \, \lambda}{2.355 \, D_{\rm dish}}\right),
\end{equation}
with $\lambda$ the observing wavelength and $D_{\rm dish}$ the diameter of the telescope.

Given $\sigma_{\rm rms}$, we compute the corresponding noise power spectrum as
\begin{equation}
P^{N} = \sigma_{\rm rms}^2 \, V_{\rm pix},
\end{equation}
where $V_{\rm pix}$ is the comoving volume subtended by a beam-sized pixel and a single frequency channel.

In principle, the finite angular resolution of the instrument implies that, the effective noise
blows up on angular scales smaller than the beam size. For all of the LIM simulations used in this work, the angular pixel size of the simulation grid, $\Omega_{L}$, is much larger than the instrumental beam area, $\Omega_{\rm pix}$, over the frequencies and redshifts of interest.

In simulations, we therefore add Gaussian random noise with RMS amplitude $\sigma_{\rm rms}$ to the pure line intensity maps. Because the simulation geometry may differ from that of the instrument in terms of angular resolution and field of view, we apply a normalization to match pixel scales. Specifically, we rescale the RMS by a factor of $\sqrt{\Omega_{\rm pix} / \Omega_{L}}$, where $\Omega_{L}$ is the angular area of a single simulation pixel. 
This prescription ensures that the variance per simulation pixel corresponds to integrating the instrumental white noise over the same solid angle.

\subsection{21-cm Observations (Interferometric)}
\label{sec:21cm_noise}

\begin{table}[t!]
\centering
\begin{tabular}{lcc}
\hline\hline
\textbf{Parameter}  & \textbf{CHORD-like} \\
\hline
Central frequency \( \nu \) (MHz)                       & 396.6 \\
Dish diameter \( D_{\rm dish} \) (m)                      & 6   \\
Antenna layout                                     & Rectangular \\
Number of antennas                                       & 512 (32$\times$16) \\
Integration time per visibility \( t_{\rm int} \) (s)     & 60  \\
Observing time per day (hr)                               & 8   \\
Bandwidth \( \Delta\nu_{\rm total} \) (MHz)               & 32  \\
Receiver temperature \( T_{\rm rec} \) (K)               & 30  \\
Beam model                                          & Gaussian \\
\hline\hline
\end{tabular}
\caption{Instrumental parameters used to model thermal noise for CHORD-like 21-cm interferometric experiment. These values are used in conjunction with the \texttt{21cmSense} code to compute the 1D thermal noise power spectrum, and subsequently the noise variance for 21-cm intensity maps used in the analysis.}
\label{tab:21cm_params}
\end{table}

The instrumental noise for 21-cm interferometric experiments arises from thermal fluctuations in visibility measurements between antenna pairs. We model this noise using the \texttt{21cmSense}\footnote{\url{https://github.com/rasg-affiliates/21cmSense}} code, which computes the sensitivity based on array configuration, observing strategy, system temperature, and frequency resolution, following the methodology of Refs.~\cite{Pober:2013jna, Liu:2019awk}. In particular, for each baseline and frequency channel, the root-mean-square (RMS) noise in visibility space at a given baseline pair \( (u,v) \) is given by
\begin{equation}
\sigma_{\rm rms}(u,v) = \frac{T_{\rm sys}}{\sqrt{N_{\rm red} \, t_{\rm obs} \, \Delta\nu}} \, \Omega_{\rm surv},
\end{equation}
where \( T_{\rm sys} \) is the system temperature, \( N_{\rm red} \) is the number of redundant baselines contributing to the mode, \( t_{\rm obs} \) is the total observing time, \( \Delta\nu \) is the spectral channel width, and \( \Omega_{\rm surv} \) is the surveyed sky area.

The \texttt{21cmSense} framework accounts for the time-dependent baseline coverage induced by Earth rotation, beam chromaticity, and instrument redundancy. It uses a Gaussian primary beam model set by the dish diameter. The code calculates the thermal noise power spectrum by summing the contributions of all baselines to the Fourier modes, and subsequently averages them cylindrically to obtain the 1D noise power spectrum \( P_N(k) \).

We apply this code to model the thermal noise for the interferometric 21-cm signal in our simulations, using representative 
\textbf{CHORD}-like instrument specifications, which are summarized in Table~\ref{tab:21cm_params}. From the resulting 1D thermal noise power spectrum, we generate a 3D Gaussian random field and add the resulting realization to our simulated 21-cm intensity cubes to obtain mock noisy observations.

\nocite{*}
\bibliography{refs}{}

\end{document}